\documentclass{article}

\usepackage[english]{babel}
\usepackage{graphicx}
\usepackage{array}
\usepackage{float}
\usepackage{placeins} 
\usepackage{amsmath} 
\usepackage{multirow}
\usepackage{booktabs}
\numberwithin{table}{section} 
\usepackage[letterpaper,top=2cm,bottom=2cm,left=3cm,right=3cm,marginparwidth=1.75cm]{geometry}

\usepackage{amsmath}
\usepackage{graphicx}
\usepackage[colorlinks=true, allcolors=blue]{hyperref}
\usepackage{amssymb}

\title{HOSVD-SR: A Physics-Based Deep Learning Framework for Super-Resolution in Fluid Dynamics}
\author{
G. Barragán\textsuperscript{1},
A. Hetherington\textsuperscript{1},
R. Abadía-Heredia\textsuperscript{1}, 
J. Garicano-Mena\textsuperscript{1}, 
S. Le Clainche\textsuperscript{1,*}
}

\date{}
\begin{document}

\maketitle
$\textsuperscript{1}$ ETSI Aeronáutica y del Espacio, Universidad Politécnica de Madrid, Plaza Cardenal Cisneros, 3, Madrid, 28040, Spain.

\begin{abstract}In this work we present a novel methodology that combines Higher Order Singular Value Decomposition (HOSVD) with Deep Learning (DL) techniques for super-resolution in computational fluid dynamics (CFD) and sparse experimental datasets. This approach, referred to as HOSVD-SR \footnote{The website of the software is available at \href{https://modelflows.github.io/modelflowsapp/}{https://modelflows.github.io/modelflowsapp/}}, integrates modal decomposition techniques with Machine Learning (ML), creating a hybrid model grounded in the underlying physics of the studied phenomena and capable of enhancing data dimensionality. The proposed methodology leverages HOSVD, a robust variant of SVD, ideal for high dimensional data, which extracts the singular values and modes (spatial and temporal) associated with each dimension of the database in tensor form, reducing noise and addressing challenges related to turbulent flows. HOSVD is employed to capture the key physical patterns from a under-resolved fluid mechanics database. Each spatial mode matrix serves as input for a decoder/autoencoder type neural network trained to increase the dimensionality of the tensor and accurately reconstruct experimental and CFD-like databases. The HOSVD-SR methodology  has been tested on both two- and three-dimensional numerical databases of flow past a circular cylinder, as well as an experimental database of circular cylinder wake flows under a turbulent flow regime. HOSVD-SR has successfully addressed both laminar and turbulent cases, outperforming a previously proposed SVD-based methodology in terms of accuracy. HOSVD-SR is physics-based, making it a robust and highly generalizable method that can also be implemented for other data generation approaches in fluid mechanics problems. 
\end{abstract}

\section{Introduction}\label{introduction}
Fluid mechanics plays a major role in many natural and technical processes. Studying fluid flows improves the understanding of vital biological functions, such as oxygen and nutrient transport in the human body, as well as natural phenomena such as sea waves, rivers, and atmospheric flows. In engineering, understanding fluid flow phenomena is essential for design, manufacturing, and optimization in industries such as aerospace, chemical, and environmental engineering. Additionally, it supports informed decision-making to address issues like air pollution in urban environments~\cite{Durst2022}.

Research on fluid mechanics-related phenomena in both industry and academia is primarily conducted through experimental and numerical methods. Designing and executing experiments involve controlling numerous parameters and ensuring the correct instrumentation of the physical setup and available approaches. As a result, experiments can often be expensive and produce noisy, unresolved databases ~\cite{diaz_morales2024,saha2020}. Numerical methods, such as Computational Fluid Dynamics (CFD), have been widely used over the last few decades to solve fluid dynamics problems. These methods represent a cost-effective advancement in understanding and predicting fluid flow behavior. However, conducting accurate CFD simulations for turbulent flows can entail significant computational expenses related to processing capacity and data storage, or even be impossible to handle for high Reynolds numbers due to the high spatio-temporal scales~\cite{vinuesa2021}. These limitations accentuate the need for developing new physics-based methodologies that can improve the accuracy of both experimental and CFD-generated data while minimizing computational costs.

The latest technological advancements in computer hardware and software, data handling and storage, and the development of powerful algorithms, coupled with the vast amount of data resulting from previous research (benchmark test cases, CFD simulations, experiments, etc.), are enabling fluid mechanics research to reach new levels~\cite{brunton2020}. Machine learning techniques are widely used by researchers among different disciplines because of their capacity to learn from data without requiring explicit mathematical models. In fluid mechanics, deep learning models are widely used because their neural network (NN) approach is well-suited for modelling physical phenomena. This method breaks down complex target functions into simpler components, which helps reduce sample time and complexity while maintaining the non-linear characteristics of the data~\cite{vinuesa2022}. In this regard, machine learning has already been used in a wide range of fluid mechanics applications such as turbulence modelling, reduced-order modelling, data reconstruction, super-resolution, and flow control.

Super-resolution began as a technique for enhancing image quality, capable of reconstructing high-resolution images from their low-resolution counterparts~\cite{LEPCHA2023230}. The implementation of ML techniques for super-resolution made possible the reconstruction of finer quality images due to its ability to find a nonlinear relationship between input and output data even under poor conditions~\cite{dar2024}. This approach has been adopted by the fluid mechanics research community to improve the accuracy of low-resolution data generated from CFD simulations or experiments, changing the “RGB” color channels for the “u, v, w” velocity fields, pressure fields, etc~\cite{SOFOS2024106396},~\cite{PANG2024106249}.
Reduced-order modelling (ROM) are high-fidelity techniques suitable for the computational complexity reduction of full-order models (FM)~\cite{volkwein2024}. ROMs are simplified representations that capture the most significant properties of a physics-based full-order model while neglecting less critical aspects. These models closely approximate the behavior of the FM~\cite{guo2022}. Thus, a ROM can be considered a low-dimensional representation of the actual FM dynamics~\cite{dar2024}. There are two types of ROMs: intrusive (iROMs) and non-intrusive (niROMs). iROMs require the underlying equations of the system being available because they involve directly modifying the governing equations to create a simplified, reduced-order version. The most common approaches are based on the Galerkin projection of full-state equations into lower-dimensional subspaces defined using a modal basis. In contrast, niROMs do not need the underlying governing equations to accurately represent the phenomenon. Several authors have successfully developed niROMs for turbulent compressible flows and transport phenomena~\cite{holemans2022,solan_fustero2023,kaneko2022,ivagnes2023}.

Non-intrusive reduced-order modelling methods are capable of representing the dynamics of a full-order model in a simpler low-dimensional system, while deep learning techniques can effectively address the non-linearity of the phenomena~\cite{fukami2023}. In this context, Abadía-Heredia et al.\cite{ABADIAHEREDIA2022115910} demonstrated the applicability of coupling reduced-order modelling techniques with machine learning under a hybrid approach (referred to as hybrid-ML) through the development of a predictive ROM capable of reducing the dimensionality of a database via proper orthogonal decomposition (POD) and predicting the future behavior of the phenomenon via convolutional neural networks. Le Clainche et al.\cite{LeClainche_Rosti_Brandt_2022}, in their work, proposed a novel technique to develop a ROM by high-order dynamic mode decomposition (HODMD) and deep learning, which is capable of predicting the wall-shear stress statistics of a turbulent flow over an isotropic porous wall. In the same line, Mata et al.\cite{MATA2023120817} developed a hybrid-ML ROM capable of forecasting the behavior of a multi-phase flow of two-concentric jets through the combination of HODMD, convolutional and recurrent neural networks. 

Díaz-Morales et al.~\cite{diaz_morales2024} proposed a singular value decomposition (SVD)-based ROM capable of enhancing the spatial resolution of fluid dynamics databases obtained from sparse data originating from experimental or coarse-grid CFD simulations. In their work, the resolution enhancement is performed through the expansion of the left and right singular vectors resulting from the SVD via a fully connected neural network consisting of two dense layers. This methodology has been applied to different fluid dynamics databases, and its applicability is limited when dealing with highly complex, noisy, and turbulent datasets.  

This article proposes, for the first time to the authors' knowledge, a hybrid-ML methodology that combines machine learning with high-order Singular Value Decomposition (HOSVD) to enhance the accuracy of coarse-grid databases derived from CFD simulations or experimental data while maintaining low dimensionality. In this work, we demonstrate how HOSVD-SR outperforms the hybrid ML ROM presented by some of the co-authors in Ref.~\cite{diaz_morales2024}, referred to as SVD-SR, showing a significant improvement in terms of accuracy, since each of the dimensional components of the tensor and noise are addressed dimension-wise. The highlights of this work include: (i) the development of a niROM based on physical principles; (ii) the combination of HOSVD with neural networks to perform resolution enhancement while keeping the reduced dimensionality; and (iii) the considerable reduction of computational time for the reconstruction of fluid dynamics data. The present methodology has been applied to databases with diverse complexity: 2D flow and 3D flow in a transient regime around a cylinder, with datasets generated by a CFD solver, as well as an experimental dataset of a turbulent cylinder wake. The robustness of the methodology makes it an optimal tool for the analysis and data reconstruction of other fluid dynamics databases.

This manuscript is organized as follows. Section ~\ref{methodology} describes the proposed niROM methodology based on the combination of HOSVD with DL strategies, referred to as HOSVD-DL throughout this work; Section \ref{Error} describes how to compute the error metrics; Section \ref{secTest} details the databases used to validate the proposed methodology;  Section \ref{results and discussion} describes the main results obtained and the discussion. Finally, Section \ref{conclusions} reports the conclusions of the work.

\section{Methodology}\label{methodology}

The following Section provides a detailed description of the steps taken to develop the proposed resolution enhancement hybrid-ROM. These are, the organization of the numerical or experimental data in matrix or tensor form, the approach utilized for the spatial compression, a brief explanation of both SVD and HOSVD modal decomposition techniques, followed up by a brief description of the the SVD hybrid machine learning model to enhance data resolution, named as SVD-SR. Finally, an in-depth description of the proposed HOSVD-based machine learning hybrid model, named as HOSVD-SR, to enhance data resolution is presented. Additionally, the error metrics considered for the evaluation of the proposed method with its counterpart are also presented.    

\subsection{Data Organization}  

Fluid dynamics databases obtained through numerical simulations with a structured or unstructured mesh are composed by the streamwise, normal and spanwise velocity components (\(u\), \(v\), and \(w\)), which are associated to the spatial dimensions (\(x\), \(y\), and \(z\)) of the fluid flow. These databases must be arranged either as a matrix or as a multidimensional array (tensor) for further analysis.

Databases organized as matrices consist of a set of \( K \) snapshots \( \boldsymbol{v}_k = \boldsymbol{v}(t_k) \),  
where \( t_k \) is the discrete time at instant \( k \).  
For convenience, these snapshots are collected in the following \emph{snapshot matrix}:
\begin{equation}
\boldsymbol{V}_1 ^ K  = [\boldsymbol{v}_1, \boldsymbol{v}_2, \dots, \boldsymbol{v}_k, \boldsymbol{v}_{k+1}, \dots, \boldsymbol{v}_{K-1}, \boldsymbol{v}_K].
\label{equation: snapshot_matrix}    
\end{equation}

For multiple variable cases, it is possible to arrange the information into multi-dimensional arrays, referred to as \emph{snapshot tensors}, where information related to flow properties (such as velocity components, pressure, etc.), spatial coordinates, and time are organized separately across different tensor components.

This snapshot tensor consists of a collection of matrix snapshots stored as a multidimensional array with more than two indexes, where the \emph{fibers} of a tensor correspond to the matrix columns and rows for a given time. In this work, the two- and three-dimensional databases where organized as fourth and fifth order tensors respectively . 

\begin{figure}[H]
    \centering
    \includegraphics[width=12 cm]{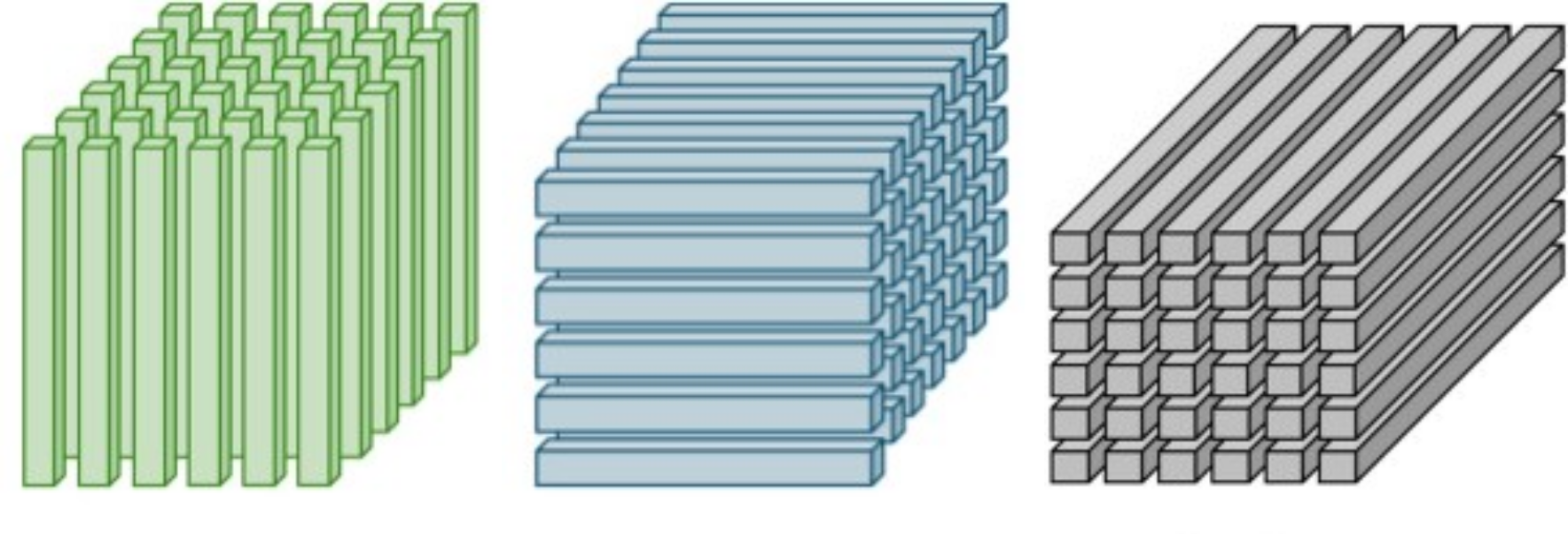} 
    \caption{Fibers of a three dimensional tensor.}
    \label{fig:fibers}
\end{figure}

For the two-dimensional time-dependent datasets described above, the streamwise and normal spatial components are represented in a \(J_2 \times J_3\) coordinate system as:

\begin{equation}
    \boldsymbol{v}(x_{j_2}, y_{j_3}, t_k) \quad \text{for } j_2 = 1, \dots, J_2, \quad j_3 = 1, \dots, J_3, \quad k = 1, \dots, K.
\end{equation}

These snapshots are re-organized in a fourth order \(J_1 \times J_2 \times J_3 \times K\)-tensor \(\mathbf{V}\), whose components \(V_{j_1 j_2 j_3 k}\) are defined as follows :

\begin{equation}
\label{equation:4dimtensor}
    V_{1 j_2 j_3 k} = v_x (x_{j_2}, y_{j_3}, t_k), \quad 
    V_{2 j_2 j_3 k} = v_y (x_{j_2}, y_{j_3}, t_k).
\end{equation}

where the index \( {j_1} \) refers to the components of the dataset, in this particular case \( {j_1} = 1,2\) where \(J_1 = 2\) for the streamwise and normal velocity components, \( {j_2} \) and \({j_3} \) are the discrete values of the two spatial coordinates,and \( k \) represents the discrete time index. It is worth mentioning that the \( {j_1} \) can take more values, depending on the fluid dynamics phenomenon under analysis. 

For three-dimensional databases utilized in this work, the streamwise, normal, and spanwise velocity components are organized in a \(J_2 \times J_3 \times J_4\) coordinate system as follows:

\begin{equation}
\begin{aligned}
\label{equation:5dimtensor}
    V_{1 j_2 j_3 j_4 k} &= v_1 (x_{j_2}, y_{j_3}, z_{j_4}, t_k), \\
    V_{2 j_2 j_3 j_4 k} &= v_2 (x_{j_2}, y_{j_3}, z_{j_4}, t_k), \\
    V_{3 j_2 j_3 j_4 k} &= v_3 (x_{j_2}, y_{j_3}, z_{j_4}, t_k), \\
\end{aligned}
\end{equation}

where the index \( {j_1} \) refers to the streamwise, normal and spanwise velocity components (\( {j_1} = 1,2,3\) where \(J_1 = 3\)), \( {j_2} \), \({j_3} \), \({j_4} \)  are the discrete values of the \(x\), \(y\) and \(z\)spatial coordinates, while \( k \) is the discrete time instant.

The snapshot matrix and snapshot (as in equations (\ref{equation:4dimtensor}) and (\ref{equation:5dimtensor})) tensor formulations are closely related and can be transformed into one another through a systematic reshaping process. In the snapshot matrix representation, the tensor indices \( j_1, j_2, j_3 \) (and \( j_4 \) for three-dimensional datasets) are collapsed into a single index \( j \). This results in a data structure where each snapshot is stored as a matrix \( V_j^k \in J \times K \), with \( J = J_1 \times J_2 \times J_3 (\times J_4) \).

\subsection{Data compression process}

\begin{figure}[H]
    \centering
    \includegraphics[width=12 cm]{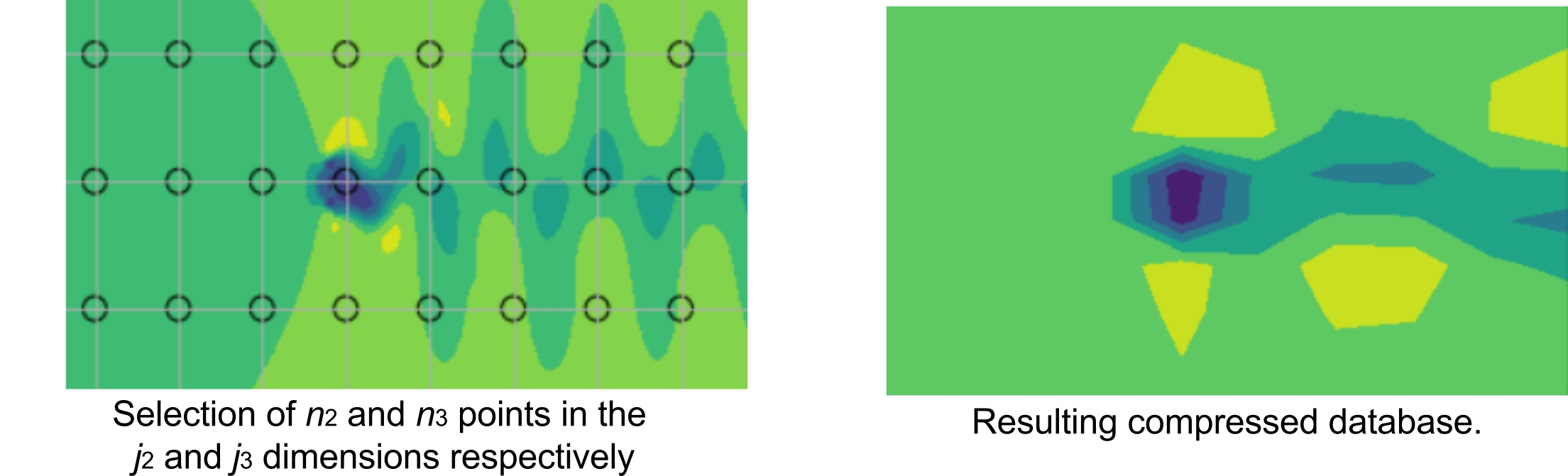} 
    \caption{Example of the compression process to model coarse meshes of two-dimensional CFD simulations or sparse sensor experimental databases, which consists on an equi-spaced point selection of the database, taking $1$ every $20$ points. The dimensions of the original and low-resolution datasets are $199$ $\times$ $499$ and $10$ $\times$ $40$ grid points, respectively. The same process can be applied for three-dimensional databases.}
    \label{fig:decim}
\end{figure}

In the case of the tensor compression process, we aim to reduce the datasets corresponding to the full-order models in terms of spatial resolution to mimic the results of coarse-grid CFD simulations or those obtained from experimental approaches. As shown in Fig.~\ref{fig:decim}, the compression process is performed by selecting every \(n\)-th point in each spatial dimension $x$, $y$ and $z$. For the case of a fourth-order tensor, we select \(n_{2}, n_{3}\) points from the \(j_{2}, j_{3}\) tensor components, which correspond to the \(x\) and \(y\) spatial coordinates, respectively. For fifth-order tensors, the selected points \(n_{2}, n_{3}, n_{4}\) correspond to the \(j_{2}, j_{3}, j_{4}\) tensor components, which represent the \(x, y\), and \(z\) spatial coordinates, respectively.

The low-resolution tensor for a four-dimensional case can be expressed as:

\begin{equation}
{\mathbf{V}_1}^{DS}(j_2, j_3, k) \quad \text{for} \quad \, j_2 = 1, \dots, N_{s2}, \, j_3 = 1, \dots, N_{s3}, \, k = 1, \dots, K.
\label{equation:tensor_4dim}
\end{equation}

\begin{equation}
\mathbf{V}_2^{DS}(j_2, j_3, k) \quad \text{for} \quad \, j_2 = 1, \dots, N_{s2}, \, j_3 = 1, \dots, N_{s3}, \, k = 1, \dots, K.
\label{equation:tensor_4dim}
\end{equation}

The resulting low-resolution tensor is denoted as \(\textbf{}{V}^{DS}_{j_1, j_2, j_3, k}\), where \(j_1 = 1, \ldots, J_{1}\) corresponds to the velocity components, \(j_2 = 1, \ldots, N_{s2}\) corresponds to the selected points in the first spatial dimension (\(x\)), \(j_3 = 1, \ldots, N_{s3}\) corresponds to the selected points in the second spatial dimension (\(y\)), where  \(N_{s2} < J_2\) and \(N_{s3} < J_3\), and \(k = 1, \ldots, K\) for tensor component associated to the time.

For a five-dimensional tensor:

\begin{equation}
{\mathbf{V}_1}^{DS}(j_2, j_3, j_4 k) \quad \text{for} \quad \, j_2 = 1, \dots, N_{s2}, \, j_3 = 1, \dots, N_{s3}, j_4 = 1, \dots, N_{s4},\, k = 1, \dots, K,
\label{equation:tensor_4dim}
\end{equation}

\begin{equation}
\mathbf{V}_2^{DS}(j_2, j_3,j_4, k) \quad \text{for} \quad \, j_2 = 1, \dots, N_{s2}, \, j_3 = 1, \dots, N_{s3}, j_4 = 1, \dots, N_{s4}, \, k = 1, \dots, K,
\label{equation:tensor_4dim}
\end{equation}

\begin{equation}
\mathbf{V}_3^{DS}(j_2, j_3, j_4, k) \quad \text{for} \quad \, j_2 = 1, \dots, N_{s2}, \, j_3 = 1, \dots, N_{s3},j_4 = 1, \dots, N_{s4}, \, k = 1, \dots, N_t.
\label{equation:tensor_4dim}
\end{equation}

The resulting low-resolution tensor is denoted as \(\textbf{}{V}^{DS}_{j_1, j_2, j_3, j_4, k}\), where \(j_1 = 1, \ldots, J_{1}\) corresponds to the velocity components, \(j_2 = 1, \ldots, N_{s2}\) corresponds to the selected points in the first spatial dimension (\(x\)), \(j_3 = 1, \ldots, N_{s3}\) corresponds to the selected points in the second spatial dimension (\(y\)), \(j_4 = 1, \ldots, N_{s4}\) corresponds to the selected points in the third spatial dimension (\(z\)), where  \(N_{s2} < J_2\), \(N_{s3} < J_3\) and \(N_{s4} < J_4\). For the temporal component of the tensor \(k = 1, \ldots, K\).

In tensor compression, the Compression Ratio (CR) compares the number of original grid points with the number of reduced grid points (or sensors, if that term is used in the document). For example, a CR of 10 (also written as 1:10) means that the compressed data uses 10 times fewer points than the original data that is, only one tenth of the original spatial information is retained.

\subsection{Singular value decomposition}

The Singular Value Decomposition (SVD) modal decomposition method was introduced as a mathematical approach for fluid mechanics applications by Sirovich \cite{Sirovich}, that is capable of extracting coherent structures from laminar and turbulent flows. SVD decomposes a matrix consisting of spatio-temporal data $\mathcal{\boldsymbol{v}}(x,y,z,t)$ into a set of proper orthogonal spatial SVD modes, also known as proper orthogonal decomposition (POD) modes \cite{10.1007/978-3-030-20055-8_53}. These can be represented as $\Phi_{n}(x,y,z)$ which are weighted by its temporal coefficients denoted as $\mathcal{\boldsymbol{c}}_n(t)$ as follows:

\begin{equation}
    \mathcal{\boldsymbol{v}}(x,y,z,t) \approx \sum_{n=1}^{N} \mathcal{\boldsymbol{c}}_n(t)\, \Phi_{n} (x,y,z).
\label{equation:svd_1}    
\end{equation}

The  \( \boldsymbol{V}_1 ^ K\) in eq. (\ref{equation: snapshot_matrix}) snapshot matrix can be decomposed using SVD into the spatial (or SVD) modes denoted as \(\boldsymbol{W}\), the temporal coefficients \(\boldsymbol{T}\) and the singular values \(\mathbf{\Sigma}\), as:
\begin{equation}
    \boldsymbol{V}_1^K = \boldsymbol{W}_{J \times J} \boldsymbol{\Sigma}_{J \times K} \boldsymbol{T}_{K \times K}^T
    \label{equation: SVD_eqn}
\end{equation}
where the columns of \(\mathbf{ W} \) and \( \mathbf{T} \) are orthonormal, and \( \mathbf{\Sigma} \) is a diagonal matrix containing the singular values \(\sigma_1, \ldots, \sigma_n \)of \( \mathbf{V}_1 ^ K \), ordered from greatest to smallest energy content, all real values. The subindexes in eq. \ref{equation: SVD_eqn} represent the dimensions  \(\mathbf{ W} \) and \( \mathbf{T} \) matrices dimensions. The descending order of the singular values is particularly useful for the fluid dynamics problems where the modes with the highest contents of energy encompasses the general dynamics of the system. These modes represent coherent structure and principal patterns of the flow, while the ones with lesser energy content could be related to noise or spatial redundancies. This arrangement of singular values is particularly useful when performing dimensionality reduction, where a low-rank approximation of \( \mathbf{V}_1 ^ K  \) can be obtained by truncating the decomposition and retaining only the first \( r' \) SVD modes associated to the largest singular values as:

\begin{equation}
   {\mathbf{V}_1 ^ K}_{r'} \approx \mathbf{W}_{J \times r'} \mathbf{\Sigma}_{r' \times r'} \mathbf{T}^T_{r' \times K } \tag{12},
\end{equation}

where \( r' < J \) and \( r' < K \), ensuring a reduced-order representation that captures the most significant features of the data. This truncated decomposition is commonly used to to calculate the POD modes, generally used to identify coherent structures in fluid flows.

\subsection{High-order singular value decomposition}

The high-order singular value decomposition (HOSVD), also known as orthogonal Tucker decomposition, was first introduced and developed by Tucker in 1966 \cite{Tuck1963a, Tucker1966, Tuck2006} and later popularized by De Lathauwer et al.\cite{DeLathauwer2000a, DeLathauwer2000b}. HOSVD has been successfully used for several applications, such as the compression and generation of aerodynamic databases ~\cite{LORENTEcomp, Lorente2009HOSVD}, for denoising and identifying patterns in fluid dynamics databases ~\cite{HETHERINGTON2024109217}, for filtering and post-processing echocardiography databases ~\cite{Wahyulaksana2023HigherOS} and the analysis and compression of gyrokinetic databases ~\cite{HATCH20124234}. This modal decomposition technique is a multi-dimensional extension of SVD. The HOSVD of a four and five dimensional tensor can be written as: 

\begin{equation}
\mathbf{V}_{j_1 j_2 j_3 k} \approx \sum_{p_1=1}^{P_1} \sum_{p_2=1}^{P_2} \sum_{p_3=1}^{P_3} \sum_{n=1}^{N} \mathbf{S}_{p_1 p_2 p_3 p_4 n} \mathbf{W}^{(1)}_{j_1 p_1} \mathbf{W}^{(2)}_{j_2 p_2} \mathbf{W}^{(3)}_{j_3 p_3} \mathbf{T}_{kn},
\label{equation: HOSVD}
\end{equation}

\begin{equation}
\mathbf{V}_{j_1 j_2 j_3 j_4 k} \approx \sum_{p_1=1}^{P_1} \sum_{p_2=1}^{P_2} \sum_{p_3=1}^{P_3} \sum_{p_4=1}^{P_4} \sum_{n=1}^{N} \mathbf{S}_{p_1 p_2 p_3 p_4 n} \mathbf{W}^{(1)}_{j_1 p_1} \mathbf{W}^{(2)}_{j_2 p_2} \mathbf{W}^{(3)}_{j_3 p_3} \mathbf{W}^{(4)}_{j_4 p_4} \mathbf{T}_{kn},
\label{equation: HOSVD5}
\end{equation}

where $\mathbf{S}_{p_1 p_2 p_3 p_4 n}$ is the core tensor, and the matrices $\mathbf{W}^{(1)}$, $\mathbf{W}^{(2)}$, $\mathbf{W}^{(3)}$, $\mathbf{W}^{(4)}$, and \(\boldsymbol{T}\) are the set of modes resulting from the decomposition. $\mathbf{W}^{(1)}$ is referred to the number of components (i.e streamwise, normal, and spanwise velocity components for the databases sued in this study) $\mathbf{W}^{(2)}$, $\mathbf{W}^{(3)}$, $\mathbf{W}^{(4)}$ mode matrices correspond to the spatial dimensions while the last $\mathbf{T}$ matrix is related to the temporal component. Each mode matrix has been obtained applying standard SVD to each \emph{fiber} of the tensor. The set of HOSVD modes are orthonormal and correspond to their eigenvectors associated with the nonzero (positive) eigenvalues, organized in a decreasing order. It is important to note that without any truncation, eq. (\ref{equation: HOSVD}) is exact. The singular values of the decomposition are denoted as:
\begin{equation}
\boldsymbol{\sigma}^{(1)}_{p_1}, \boldsymbol{\sigma}^{(2)}_{p_2}, \boldsymbol{\sigma}^{(3)}_{p_3}, \boldsymbol{\sigma}^{(4)}_{p_4}, \boldsymbol{\sigma}^{(t)}_{n},
\label{HOSVD_modes}
\end{equation}

for $p_1$ =1,...,$P_1$, $p_2$=1,...,$P_2$, , $p_3$=1,...,$P_3$, , $p_4$=1,...,$P_4$, n= $1$,...,N, being $P_1$, $P_2$, $P_3$, $P_4$ and N the retained singular values. The truncation of the mode matrices in the HOSVD provides an approximate decomposition of the original tensor, which can be useful for noise filtering and compression tasks. After the truncation, the HOSVD eq. (\ref{equation: HOSVD}) can be written as: 

\begin{equation}
\mathbf{V}_{j_1 j_2 j_3 j_4 k} \approx \sum_{n=1}^{N} \mathbf{W}_{j_1 j_2 j_3 j_4 n} \hat{\mathbf{V}}_{kn},
\end{equation}

where N refers to the spatial complexity of the tensor which determines the number of spatial modes  $\boldsymbol{W}_{j_1 j_2 j_3 j_4 n}$ and the rescaled temporal modes \(\boldsymbol{\hat{V}}_{kn}\), the spatial modes are defined as follows:  

\begin{equation}
\mathbf{W}_{j_1 j_2 j_3 j_4 n} = \sum_{p_1=1}^{P_1} \sum_{p_2=1}^{P_2} \sum_{p_3=1}^{P_3}  \mathbf{S}_{p_1 p_2 p_3 n} \mathbf{W}^{(1)}_{j_1 p_1} \mathbf{W}^{(2)}_{j_2 p_2} \mathbf{W}^{(3)}_{j_3 p_3}/{\mathbf{\sigma}^{t}_{n}}, 
\end{equation}

\begin{equation}
\hat{\mathbf{T}}_{kn} = \mathbf{\sigma}^{t}_{n} \mathbf{T}_{kn},
\end{equation}

This approach aims to extract the most coherent, linearly-independent features and use them to model the flow dynamics in terms of a few modes. By doing so, it is possible to produce an accurate and compressed approximation of a fluid dynamics database \cite{Clainche_2018a}. Vega $\&$ Le Clainche \cite{Clainche_2018b} illustrated the applicability of HOSVD in fluid dynamics problems, where the most significant patterns of a flow are stored in mode matrices, sorted in descending order regarding its energy content. These methods can retain the main physics of the input data, making this approach physics-based.

The HOSVD algorithm implemented in this work, contain three steps unfolding the tensor, perform SVD to each unfolded matrices and finally folding the matrices again. Each one of these steps is detailed as follows:  

Unfolding, also known as \emph{matricization}, is the process of reshaping a higher-order tensor into a matrix by reorganizing its elements along a specific mode. Given the five-dimensional tensor \( \boldsymbol{V}_{j_1 j_2 j_3 j_4 k} \), the unfolding along each mode is defined as follows:

\begin{equation}
\boldsymbol{V}_{(1)} \in \mathbb{R}^{J_1 \times (J_2 J_3 J_4 K)},
\end{equation}

\begin{equation}
\boldsymbol{V}_{(2)} \in \mathbb{R}^{J_2 \times (J_1 J_3 J_4 K)},
\end{equation}

\[
\vdots
\]

\begin{equation}
\boldsymbol{V}_{(4)} \in \mathbb{R}^{J_4 \times (J_1 J_2 J_3 K)},
\end{equation}

\begin{equation}
\boldsymbol{V}_{(5)} \in \mathbb{R}^{K \times (J_1 J_2 J_3 J_4)}.
\end{equation}

where \( \boldsymbol{V}_{(1)} \) represents the tensor unfolded along the first spatial dimension, \( \boldsymbol{V}_{(2)} \), \( \boldsymbol{V}_{(3)} \), and \( \boldsymbol{V}_{(4)} \) correspond to the other spatial dimensions, and \( \boldsymbol{V}_{(5)} \) represents the temporal unfolding. Each unfolded matrix \( \boldsymbol{V}_{(j)} \) is then decomposed using SVD:

\begin{equation}
\boldsymbol{V}_{(1)} = \boldsymbol{W}^{(1)} \boldsymbol{\Sigma}^{(1)} \boldsymbol{Q}^{(1)T},
\end{equation}

\begin{equation}
\boldsymbol{V}_{(2)} = \boldsymbol{W}^{(2)} \boldsymbol{\Sigma}^{(2)} \boldsymbol{Q}^{(2)T},
\end{equation}

\[
\vdots
\]

\begin{equation}
\boldsymbol{V}_{(5)} = \boldsymbol{T} \boldsymbol{\Sigma}^{(5)} \boldsymbol{Q}^{(5)T},
\end{equation}

where \( \boldsymbol{W}^{(j)} \) contains the orthonormal mode matrices for the spatial dimensions, \( \boldsymbol{\Sigma}^{(j)} \) is a diagonal matrix containing singular values, and \( \boldsymbol{Q}^{(j)} \) consists of the right singular vectors. The matrix \( \boldsymbol{T} \) is the temporal mode matrix obtained from the unfolding along the temporal dimension. The singular values of each unfolded matrix are stored in a list, denoted as \( \boldsymbol{\Sigma}^{(j)} \).

The core tensor \( \boldsymbol{S} \) is computed as:

\begin{equation}
\boldsymbol{S} = \boldsymbol{V} \times \boldsymbol{T}^T \times \boldsymbol{W}^{(4)T} \times \boldsymbol{W}^{(3)T} \times \boldsymbol{W}^{(2)T} \times \boldsymbol{W}^{(1)T}
\label{eq:HOSVD_reconstruction}
\end{equation}

where \(\mathbf{V}\) is the snapshot tensor. Folding, or \emph{tensorization}, is the inverse operation, where the unfolded representation is reshaped back into the original multi-dimensional tensor. Given the core tensor \( \boldsymbol{S} \) and mode matrices \( \boldsymbol{W}^{(j)} \), the approximation of \( \boldsymbol{V} \) is given by:

\begin{equation}
\label{equation: folding}
\boldsymbol{V} \approx \boldsymbol{S} \times \boldsymbol{W}^{(1)} \times \boldsymbol{W}^{(2)} \times \boldsymbol{W}^{(3)} \times \boldsymbol{W}^{(4)} \times \boldsymbol{T}.
\end{equation}

After truncation, the tensor can be approximated as:

\begin{equation}
\boldsymbol{V}_{j_1 j_2 j_3 j_4 k} \approx \sum_{p_1=1}^{P_1} \sum_{p_2=1}^{P_2} \sum_{p_3=1}^{P_3} \sum_{p_4=1}^{P_4} \sum_{k'=1}^{K'} \boldsymbol{W}_{j_1 j_2 j_3 j_4 k'} \, \hat{\boldsymbol{T}}_{k k'}.
\end{equation}

where \( \boldsymbol{W}_{j_1 j_2 j_3 j_4 k'} \) are the spatial modes, and \( \hat{\boldsymbol{T}}_{k k'} \) represents the rescaled temporal modes. The spatial modes are then defined as:

\begin{equation}
\boldsymbol{W}_{j_1 j_2 j_3 j_4 k'} = \sum_{p_1=1}^{P_1} \sum_{p_2=1}^{P_2} \sum_{p_3=1}^{P_3} \sum_{p_4=1}^{P_4} \boldsymbol{S}_{p_1 p_2 p_3 p_4 k'} \boldsymbol{W}_{j_1 p_1}^{(1)} \boldsymbol{W}_{j_2 p_2}^{(2)} \boldsymbol{W}_{j_3 p_3}^{(3)} \boldsymbol{W}_{j_4 p_4}^{(4)} / \sigma_{k'}^{(5)},
\end{equation}

and the rescaled temporal modes are given by:

\begin{equation}
\hat{\boldsymbol{T}}_{k k'} = \sigma_{k'}^{(5)} \boldsymbol{T}_{k k'}.
\end{equation}

HOSVD is applied to the low-resolution tensor, retaining an appropriate number of modes for each spatial dimension to reconstruct a simplified, compressed, and de-noised version of the original dataset. The resulting factor or mode matrices for the spatial dimensions \( \mathbf{W}^{(1)} \), \( \mathbf{W}^{(2)} \), \( \mathbf{W}^{(3)} \), and \( \mathbf{W}^{(4)} \) have sizes \( n_1 \times m_1 \), \( n_2 \times m_2 \), \( n_3 \times m_3 \), and \( n_4 \times m_4 \), respectively, where \( n_i \) corresponds to the number of elements in the low-resolution tensor along spatial dimension \( i \), and \( m_i \) denotes the number of retained SVD modes in that dimension. For each matrix \( \boldsymbol{W}^{(i)} \), the row index \( j_i = 1, \dots, n_i \) corresponds to positions in physical space (e.g., spatial grid points), while the column index \( p_i = 1, \dots, m_i \) indexes the orthogonal basis functions (modes) along that direction. These matrices and singular values serve as inputs for the neural network (NN) described in the following section. The tensor reconstruction is achieved by sequentially multiplying the core tensor \( \mathbf{S} \) by each factor matrix along the corresponding modes.

\subsection{Singular value decomposition superresolution model}

The SVD superresolution (SVD-SR) hybrid machine learning model was first introduced in Ref. ~\cite{diaz_morales2024}. This methodology leverages the decomposition capability of SVD to extract the most representative flow patterns by merging each of the decomposed  matrices \( \boldsymbol{W} \), \( \boldsymbol{T} \)  and \( \boldsymbol{\Sigma} \) with a NN. 
In their work, the decomposition is performed by applying the \emph{numpy} \cite{harris2020} implementation of the SVD available in \emph{python} on each low-resolution snapshot $k$. The \( \mathbf{W} \), \( \mathbf{T} \)  and \( \mathbf{\Sigma }\) components resulting from the decomposition are stored as tensor form and for simplicity renamed as: 

\begin{equation}
\mathbf{V}_{W}^{DS} = \mathbf{W}_{k} 
\end{equation}

\begin{equation}
\mathbf{V}_{\Sigma}^{DS} = \mathbf{\Sigma}_{k} 
\end{equation}

\begin{equation}
\mathbf{V}_{T}^{DS} = \mathbf{T}_{k} 
\end{equation}

\begin{equation}
\text{for } k = 1, \dots, K.
\end{equation}

\begin{figure}[H]
    \centering
    \includegraphics[width=0.5\linewidth]{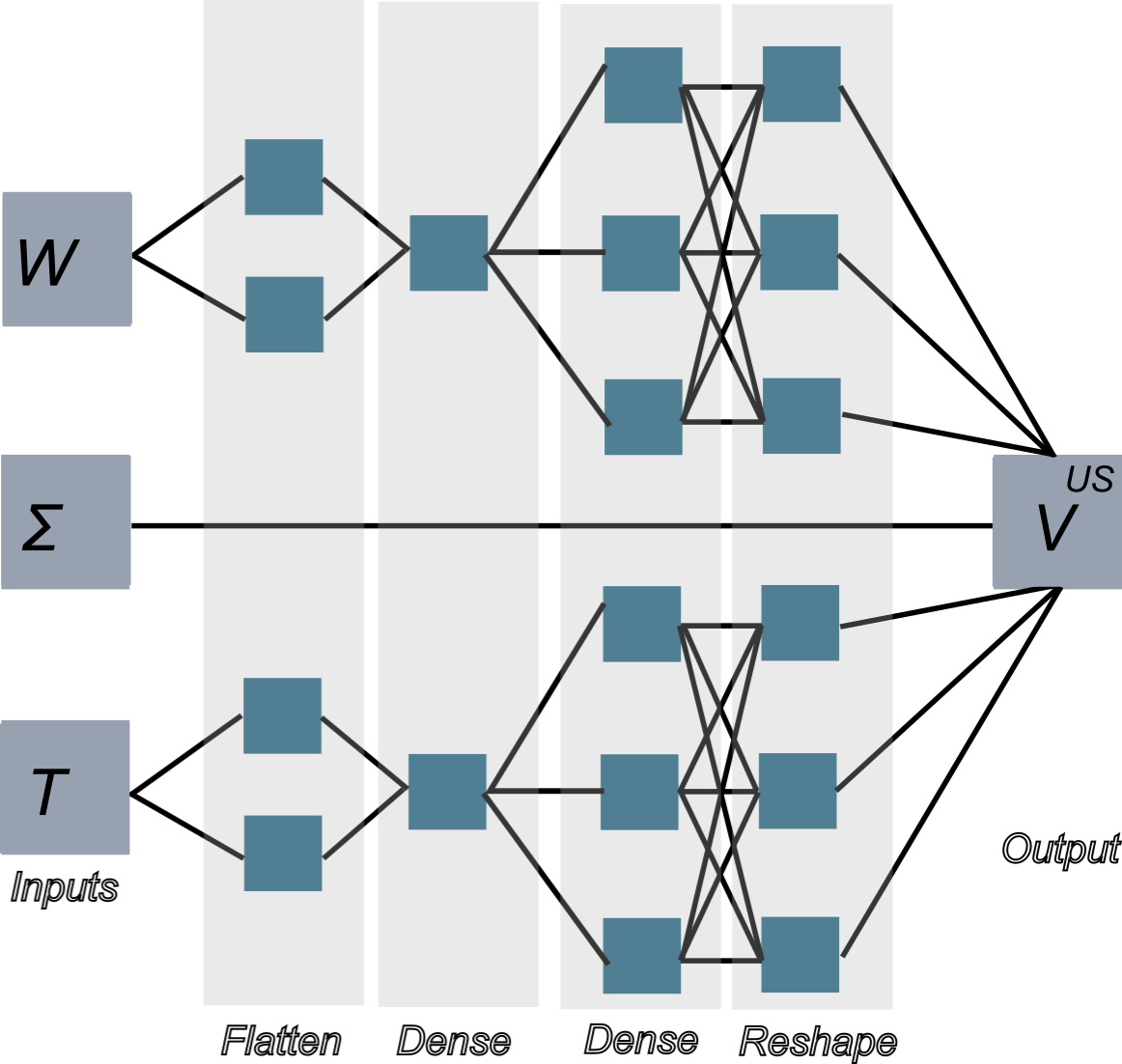} 
    \caption{Deep learning architecture of the SVD-SR approach in Ref. ~\cite{diaz_morales2024}}
    \label{fig:svdsr}
\end{figure}

The resulting tensors \( \mathbf{W}_{k} \), \( \mathbf{V}_{k} \), and \( \mathbf{\Sigma}_{k} \) serve as inputs for a neural network with an auto encoder-like architecture, which consists of three steps: encoding, decoding, and reconstruction. The encoding step is applied to the \( \mathbf{W}_{k} \) and \( \mathbf{V}_{k} \) tensors and consists of a ``flatten'' layer followed by a ``dense'' layer with a reduced number of neurons, designed to search for a low-dimensional latent space that captures the most important features necessary to learn and represent the data. The decoding step consists of a dense layer with a number of neurons equal to the final, high dimensional, shape of \( \mathbf{W}_{k} \) and \( \mathbf{T}_{k} \), followed by a ``reshape'' layer. In the reconstruction step, the reshaped \( \mathbf{W}_{k} \) and \( \mathbf{T}_{k} \) tensors are combined with \( \mathbf{\Sigma}_{k} \), which remains unchanged, and finally, the up scaled matrix \( \mathbf{V}^{US} \) , which represents the reconstruction of the snapshot matrix eqs. (\ref{equation:4dimtensor} and \ref{equation:5dimtensor})  with dimensions \(J \times K\), is obtained. This final reconstruction is then compared with the ground truth data for training the network, as shown in Fig.~\ref{fig:svdsr}. 

As in typical machine learning workflows, data is partitioned into three separate groups, with 70\% allocated for model training, 15\% reserved for validation, and the remaining 15\% set aside for performance evaluation on unseen data.  The amount of snapshots for each of the databases tested using HOSVD-SR is detailed in Tab. \ref{tab:trainvalidationtest}.

\subsection{High-order singular value decomposition superresolution model}

\begin{figure}[H]
    \centering
    \includegraphics[width=\textwidth]{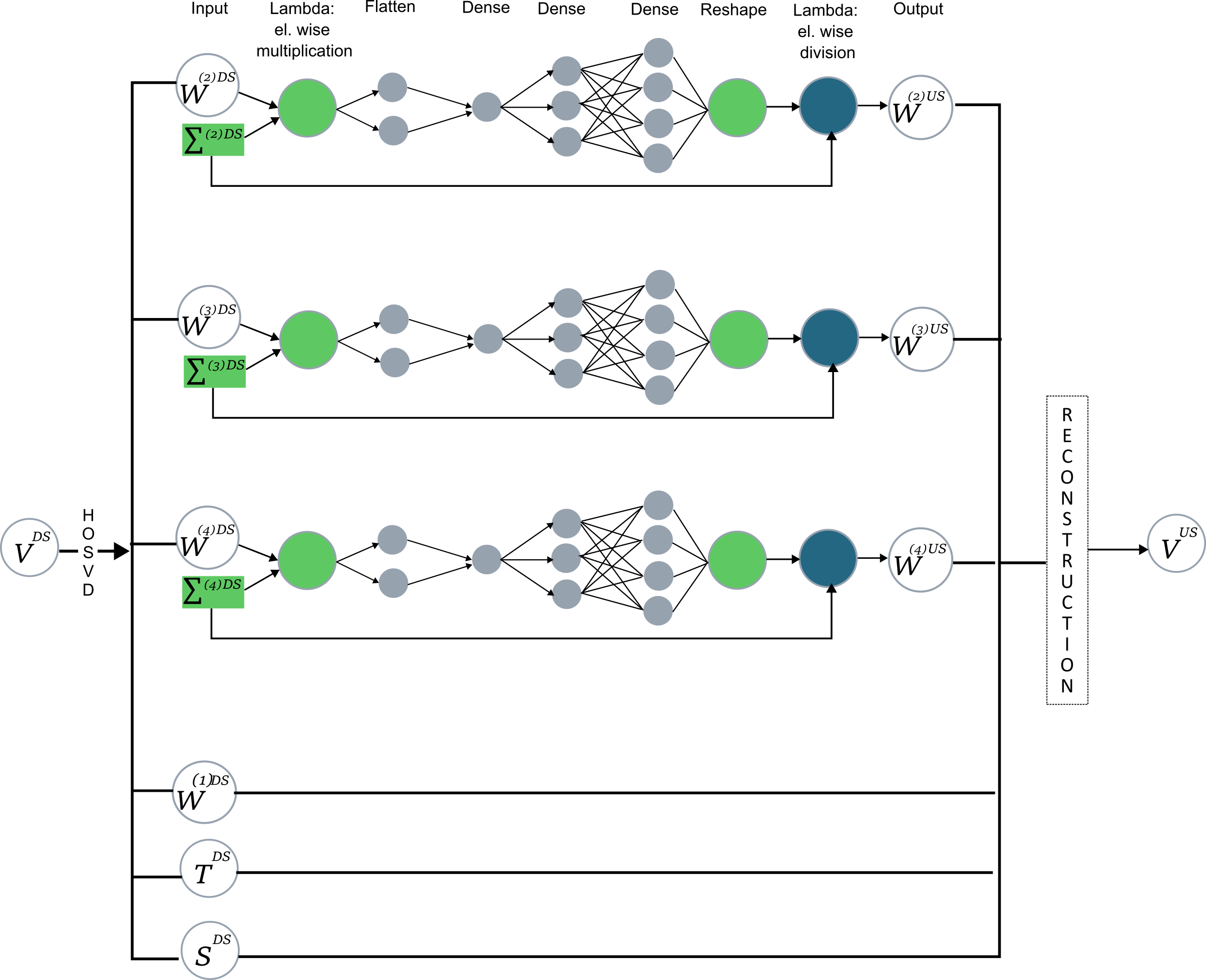} 
    \caption{Deep learning architecture for the proposed HOSVD-SR resolution enhancement tool. The low-resolution tensor first is decomposed using HOSVD -eq.~\ref{equation: HOSVD}-, then the spatial mode matrices and their singular values serve as input for the neural network which consists of 5 layers. Finally, the enhanced mode matrix is used to reconstruct the resolution-enhanced tensor.   }
    \label{fig:dl}
\end{figure}

The proposed high-order SVD superresolution (HOSVD-SR) model employs a meta-learning approach, leveraging the HOSVD's capability to extract the most significant flow patterns and structures from a fluid dynamics database in tensor form.This is an advantage, as the data can be treated independently as a function of spatial direction and time, improving clarity, removing spatial redundancies and noise, and identifying relevant patterns associated with each component of the tensor. Each mode matrix, corresponding to the spatial dimensions resulting from the HOSVD operation, is fed into an\emph{independent} neural network responsible for resolution enhancement. These mode matrices serve as the input to a neural network, referred in this work as ``reconstruction neural network'', where the tensor reconstruction and learning stage are performed. Figure \ref{fig:dl} shows the deep learning approach developed in this work. 

Each of the above-mentioned independent neural networks employs a fully connected decoder or autoencoder architecture, consisting of 5 layers: an input layer, 3 hidden layers, and an output layer. The input layer is fed by one of the low-resolution spatial factor matrices \( \mathbf{W}^{(2)DS} \), \( \mathbf{W}^{(3)DS} \), \( \mathbf{W}^{(4)DS} \) and their corresponding singular value list \( \mathbf{\Sigma}^{(2) DS} \), \( \mathbf{\Sigma}^{(3)DS} \) and \( \mathbf{\Sigma}^{(4)DS} \). The first hidden layer performs an element-wise multiplication between the mode matrix values \( \mathbf{W}^{(2,3,4)DS} \) and their corresponding singular values \( \mathbf{\Sigma}^{(2,3,4)DS} \) executed in a lambda layer, this provides initial weights to the matrices and ease the training process. A  ``flatten'' layer then converts the matrix from the previous layer into a vector. This vector is subsequently processed through 3 dense layers with the \emph{rectified linear unit (ReLu)} activation function on each, which introduces the property of nonlinearity to the model, representing a simplified decoder or autoencoder. It is important to ensure that the number of neurons in the final layer matches the desired high-resolution shape. The last hidden layer reshapes the vector into a matrix of the specified high-quality spatial mode matrix dimensions. Therefore, a element-wise division is performed by \( \mathbf{\Sigma}^{(2,3,4)} \)  which is also the output.

The reconstruction neural network combines the resolution-enhanced spatial mode matrices, called as \( \mathbf{W}^{(2) \, \text{US}} \), \( \mathbf{W}^{(3) \, \text{US}} \), \( \mathbf{W}^{(4) \, \text{US}} \), with the remaining original matrices \( \mathbf{W}^{(1)}\), $\boldsymbol{T}$ and, together with the core tensor $\mathbf{S}$, reconstructs the tensor $\mathbf{V}^{US}$ with the desired high-resolution  shape through the unfolding step in eq. (\ref{equation: folding}). The resulting tensor is compared during the training process with the high-resolution tensor $\mathbf{V}$. 

This deep learning model has been developed using the Adaptive Moment Estimation Algorithm (Adam) ~\cite{kingma2015adam}, which applies a stochastic gradient descent method based on the adaptive estimation of first-order and second-order moments to determine the weights and biases. This optimizer has demonstrated its robustness and efficiency in different tasks, such as deep neural networks, computer vision and natural language processing. Its main advantage among other optimizers relies on the bias correction, adaptative learning ratio and the implementation of ``momentum''to track of an exponentially decaying average of past gradients.

This architecture has been developed using \emph{Keras} \cite{chollet2015keras}, a deep learning library available in Python \cite{van1995python}, and executed within the \emph{TensorFlow} framework \cite{tensorflow2015-whitepaper}. To optimize the training process, the BayesianOptimization tuner in Keras is used to find the optimal number of neurons and learning rate. As common practice in machine learning development, both low-resolution and high-resolution tensors have been split into three different sets: training, validation, and testing, in order to determine the network's generalization capabilities with new data. The loss functions used for the training and validation process are the mean squared error (MSE) and the mean absolute error (MAE), as defined in the following section. Both metrics are crucial for assessing model accuracy. A summary of the hyperparameters used during the training of the neural network are detailed in Tab.~\ref{tab:hyperparamteres}. 

As is common practice in machine learning, the dataset has been divided into three distinct subsets: 70\% for training, 15\% for validation, and 15\% for testing. The amount of snapshots for each of the databases tested using HOSVD-SR is detailed in Tab. \ref{tab:trainvalidationtest}

\begin{table}[H]
\centering
\caption{Values of the relevant hyperparameters applied to different databases. The neurons for each layer represent the range for the optimizer, \(P_2,\space P_3,\space P_4, \) corresponds to the spatial retained SVD modes and \( J_2,\space J_3,\space J_4\) the high-resolution number of points on each spatial dimension.}
\label{tab:hyperparamteres}
\begin{tabular}{lccc}
Hyperparameter & Cyl 2D & Cyl 3D & Experimental cyl \\
\hline
neurons(Dense 1 - ``ReLu'')        & 64 -128  & 64 - 128    & 64 - 128  \\
neurons(Dense 2 - ``ReLu'')       & 64 -128  & 64 - 128    & 64 - 128  \\
neurons(Dense 3 -``linear'')       & \(P_2,\times J_2,\space P_3 \times J_3,\space P_4\times J_4\)  & \(P_2,\times J_2,\space P_3 \times J_3,\space P_4\times J_4\) & \(P_2,\times J_2,\space P_3 \times J_3,\space P_4\times J_4\)  \\
Batch size        & 16  & 64    & 64  \\
Loss Function (training)        & MSE  & MSE    & MSE  \\
Loss Function (validation) & MAE      & MAE  & MAE   \\
Learning rate  & 0.001  & 0.0005  & 0.0005   \\
\hline
\end{tabular}
\end{table}

\section{Error analysis and validation of the model} \label{Error}

In machine learning, the mean squared error (MSE) and mean absolute error (MAE) are two commonly used metrics to evaluate the performance of deep learning models. The MSE is defined as the average of the squared differences between the actual values ($\mathbf{V}_i$) and the predicted values ($\mathbf{V}^{US}_i$), which can be expressed as:

\begin{equation}
\text{MSE} = \frac{1}{K} \sum_{i=1}^{K} (\mathbf{V}_i - \mathbf{V}_i^{US})^2,
\end{equation}

where \( J \) represents the number of data points. The MSE squares the difference between the actual values \( \mathbf{V}_i \) and the predicted values \( \mathbf{V}_i^{US} \), which means it heavily penalizes large errors, making it particularly useful when large errors are undesirable in the model~\cite{Terven2023}.

On the other hand, the Mean Absolute Error (MAE) is calculated as the average of the absolute differences between the actual values ($\mathbf{V}_i$) and the predicted values (${V}_i^{US}$):

\begin{equation}
\text{MAE} = \frac{1}{K} \sum_{i=1}^{K} |\mathbf{V}_i - \mathbf{V}^{US}_i|
\end{equation}

The MAE is a linear score, meaning that all individual differences are weighted equally in the average, making it more interpretable and robust to outliers compared to MSE~\cite{Terven2023}. 

By using both, the loss function can take advantage of the robustness to outliers from MAE and penalize large errors with the MSE.

The Relative Root Mean Square Error (RRMSE) is used as a measure of how the reconstruction of the tensor \( \mathbf{V}^{US} \) approximates the real tensor \( \mathbf{V} \). Lower values of RRMSE are referred to better accuracy of the reconstruction. It is defined as:

\begin{equation}
\text{RRMSE} = \sqrt{\frac{\sum_{i=1}^{K} (\mathbf{V}_i - \mathbf{V}^{US}_i)^2}{\sum_{i=1}^{K} \mathbf{V}_i^2}},
\end{equation}

where \( \mathbf{V}_i \) represents the the real tensors \( \mathbf{V} \) and \( \mathbf{V}^{US}_i \) represents  the reconstructed tensors \( \mathbf{V}^{US} \).

The normalized error is calculated as the difference between the actual high-resolution tensor $\boldsymbol{V}$ and the reconstructed one $\boldsymbol{V}^{US}$,  normalized by the maximum absolute error to facilitate comparison across different scales. The formula for the normalized error is given by:

\begin{equation}
\epsilon_{i} = \frac{\mathbf{V}_i - \mathbf{V}^{US}_i}{\max(|\mathbf{V}_i - \mathbf{V}^{US}_i|)},
\end{equation}

where \( \mathbf{V}_i \) represents the actual value, \( \mathbf{V}^{US}_i \) represents the predicted value, and \( \max(|\mathbf{V}_i - \mathbf{V}^{US}_i|) \) denotes the maximum absolute error among all data points.

This normalization scales the error values so that the largest absolute error is scaled to 1, which facilitates comparison of errors across different datasets or model configurations. Plotting the normalized error for each point in the spatial dimension of the tensor helps to identify the locations of maximum errors and potential causes.

To analyse the distribution of reconstruction errors, the normalized error values are first flattened into a one-dimensional array. These values are then represented in a histogram, with the data normalized such that the total area under the curve is equal to one, ensuring it satisfies the properties of an estimated probability density function (PDF), expressed as:

\begin{equation} \int_{-\infty}^{\infty} \boldsymbol{f}(x)dx = 1, \end{equation}

where \(\boldsymbol{f}(x)\) is referred to the probability density function (PDF). This histogram-based analysis provides insight into whether errors are uniformly distributed or concentrated within specific ranges ~\cite{hetherington2023lowcostsingularvaluedecomposition}. Understanding these distributions is essential for evaluating the model’s performance and pinpointing areas for improvement. The number of bins in the plot is 50, where each of them represent a probability of error between 0\% to 2\%. 

\section{Test cases}\label{secTest}

The accuracy and robustness of HOSVD-SR have been demonstrated using three databases that include two- and three-dimensional numerical simulations of a laminar flow over a circular cylinder, and a two-dimensional database over a circular cylinder at a turbulent regime. These databases represent fluid dynamics phenomena. Thus, the governing equations for an incompressible, \( n \)-dimensional flow are given by the Navier-Stokes equations in conservative form. These include the conservation of mass and momentum equations: 

\begin{equation}
\nabla \cdot \mathbf{v} = 0, \label{eq:NS1}
\end{equation}

\begin{equation}
\frac{\partial \mathbf{v}}{\partial t} + (\mathbf{v} \cdot \nabla) \mathbf{v} = -\nabla p + \frac{1}{Re} \nabla^2 \mathbf{v},  \label{eq:NS1} 
\end{equation}

where \( \mathbf{v} \) is the non-dimensional velocity vector for the streamwise, normal and spanwise velocity components defined as \( (\textbf{v}_x, \textbf{v}_y, \textbf{v}_z) \), \( p \) is the pressure, Re is the Reynolds number defined as Re$=\frac{UL}{\nu}$, U and L correspond to the characteristic velocity and length, while $\nu$ is the kinematic viscosity.

\subsection{Flow past a circular cylinder}

The flow around a circular cylinder is a classical problem in fluid dynamics that serves as a benchmark for validating computational methods. At low Reynolds numbers, the flow remains steady, but as \(\text{Re}\) increases, the flow exhibits complex behavior. At \(\text{Re} \approx 49\), a Hopf bifurcation leads to an unsteady flow characterized by a von Kármán vortex street~\cite{Jackson198723}. A second bifurcation occurs at approximately Re = 189, triggering the transition from two-dimensional oscillations to a three-dimensional flow field. ~\cite{BLACKBURN_MARQUES_LOPEZ_2005}.

The von Kármán vortex street is a repeating pattern of vortices that appears in the wake of a bluff body that is represented by the vortex shedding \cite{Barkley_Henderson_1996}. The vortex shedding behind a circular cylinder occurs in a wide range of Reynolds number starting from  \( Re \sim 49 \). The origin and development of vortex shedding are closely linked to the boundary layer on a circular cylinder's surface. This boundary layer contains significant vorticity, which feeds into shear layers downstream of the separation point on both sides of the cylinder. This continuous input causes the shear layers to roll up into vortices. The two vortices each one at one side of the circular cylinder, exhibiting opposite vorticity, form an unstable pair under small disturbances. As a result, one vortex grows larger, drawing the opposing vortex across the wake. The larger vortex cuts off vorticity supply to the smaller vortex from its boundary layer as it approaches the opposite shear layer, leading to its separation and shedding of the vortex. Thus, a new vortex forms on the same side of the cylinder. This process repeats for the opposite vortex, resulting in a continuous cycle of vortex shedding \cite{Gerrard_1966}.

\subsubsection{Numerical databases}

The numerical databases tested in this study were part of previous works. The  two-dimensional (2D) cylinder database was obtained by Vega \& Le Clainche in \cite{Vega2020}  and the three-dimensional (3D) cylinder database was obtained by Le Clainche et al. in Ref ~\cite{Clainche_2018a, Clainche_2018b}. The simulations for both 2D and 3D approaches were carried out using the open-source solver Nek5000, that uses spectral elements as spatial discretization, to solve the Navier-Stokes equations under defined boundary conditions.  For both cases, the domains are discretized into rectangular elements, with polynomial order \(\mathcal{P} = 9\). The computational domains are non-dimensionalized by the cylinder's diameter, extending \(\pm 15D\) in the normal direction and \(15D\) upstream and \(50D\) downstream of the cylinder. 

\begin{figure}[H]
    \centering
    \includegraphics[width=\textwidth]{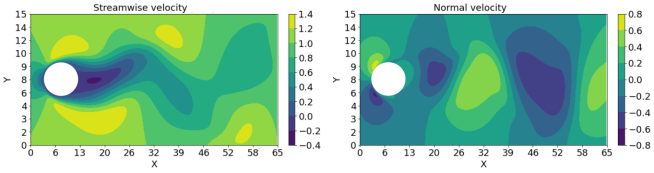} 
    \caption{Streamwise and normal  velocity components of a representative snapshot of the two-dimensional wake behind a circular cylinder at Re = 100.}
    \label{fig:example2d}
\end{figure}

The two-Dimensional Cylinder database represents the saturated flow around a 2D cylinder with a Reynolds number of 100, a representative snapshot is presented in Fig. ~\ref{fig:example2d}. This database comprises 150 time equi-spaced snapshots with a time step $\Delta$t = 0.2. Each snapshot includes two components of the velocity vector (\(v_x\), \(v_y\)). The initial conditions for this simulation are \(v_x =1, \space v_y = 0\), while the boundary conditions at the inlet were Dirichlet for velocity (with components \(v_x =1, v_y = 0\)) and at the top, bottom and output walls of the domain are Neumann for velocity and Dirichlet for pressure (p=0).

\begin{figure}[H]
    \centering
    \includegraphics[width=\textwidth]{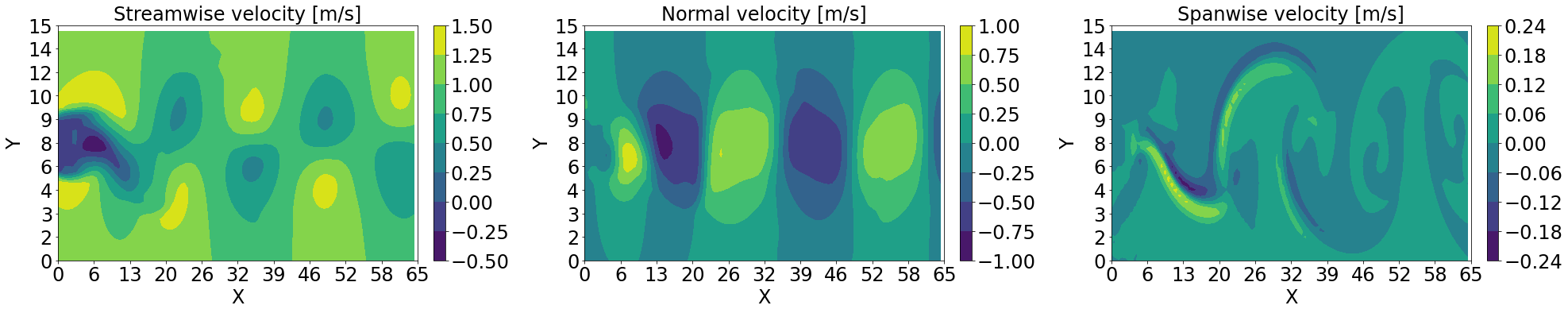} 
    \caption{streamwise, normal and spanwise velocity components of a representative snapshot of the three-dimensional wake behind a circular cylinder at Re = 280.}
    \label{fig:example3D}
\end{figure}

The three-dimensional cylinder database at \(\text{Re} = 280\) is composed by 599 snapshots that were collected from the start of the simulation with a time step \(\Delta t = 1\). However, the spanwise velocity develops only after time \(t = 350\), and the flow reaches a saturated regime by \(t = 500\)~\cite{Clainche_2018b}. Consequently, the final 299 snapshots are used to represent the saturated flow regime, a representative snapshot is presente in Fig. \ref{fig:example3D}. This database includes the three components of the velocity vector (\(v_x, v_y, v_z\)), which correspond to the streamwise, normal, and spanwise velocity components, respectively. 

Both the two- and three-dimensional numerical databases have been post processed in order to extract the area of interest, which in this case corresponds to the wake downstream the cylinder. The shapes of these databases in tensor form are detailed in Tab. \ref{table: database_summary}.

\subsubsection{Experimental database}

\begin{figure}[H]
    \centering
    \includegraphics[width=\textwidth]{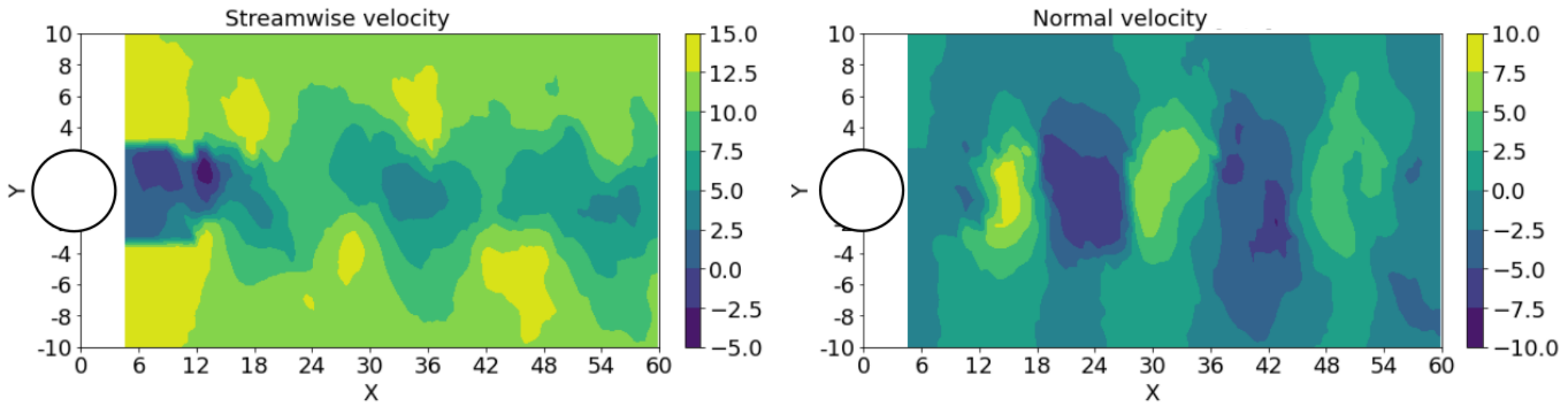} 
    \caption{streamwise (left) and normal (right) velocity components of a representative snapshot of experimental cylinder with Re = 2600 dataset from ~\cite{Mendez_2020}}
    \label{fig:vki}
\end{figure}

The experimental flow past a circular cylinder database models the three-dimensional wake behind a circular cylinder in a turbulent flow regime with $\text{Re} = 2600$. This database was obtained at the von Kármán Institute using the L10 low-speed wind tunnel, as described by \cite{Mendez_2020}. The data acquisition was performed using Particle Image Velocimetry (PIV) techniques, employing a TR-PIV system from Dantec Dynamics and a high-resolution SpeedSense 9090 camera. The L10 wind tunnel was equipped with a piezoresistive pressure transducer and a Laskin Nozzle for seeding particle generation, ensuring precise measurements under transient flow conditions around a cylinder with a diameter of $D = 5$ mm and a length of $L = 200$ mm. The field of view covered approximately $70 \times 26$ mm, with a resolution of 18.3 px/mm. The dataset consists of  $ 4000 $ snapshots as double-frame images, sampled at a frequency of $f_s = 3$ kHz. The time step is computed as $\Delta t = \frac{1}{f_s}$, yielding $\Delta t = 0.333$ \emph{ms}. Each snapshot in this dataset contains the streamwise and normal components of the velocity vector, denoted as $(v_x, v_y)$. A representative snapshot of this database is presented in Fig. \ref{fig:vki}.

\subsection{Summary of the databases}

Table~\ref{tab:sum_datasets} describes the dimensions of each of the test case databases mentioned above. For both the experimental database and the two-dimensional flow past a circular cylinder database, the two variables used in the reconstruction correspond to the streamwise and normal components of the velocity vector $(v_x, v_y)$. For the three-dimensional flow past a circular cylinder numerical database, the three variables involved are the streamwise and normal velocity components $(v_x, v_y, v_z)$.  

\begin{table}[h!]
\centering
\caption{Summary of the dimensions of the databases. The experimental database of a circular cylinder wake flows is referred as ``Experimental cyl'', the two- and three-dimensional flow past a circular cylinder databases are referred as ``Cyl 2D'' and ``Cyl 3D'' respectively. \( J_{1} \) refers to the number of variables, such as streamwise, normal and spanwise velocity components for the ``Cyl 3D'' database and streamwise and normal velocity components for the ``Experimental cyl'' and ``Cyl 2D'' databases. \( J_2 \), \( J_3 \), and \( J_4 \) represent the number of points in the \(x\), \(y\), and \(z\) space dimensions, and \( t_K \) represents the number of available snapshots.}
\label{tab:sum_datasets}
\begin{tabular}{lccccc}
\label{table: database_summary}

\textbf{Database} & \(J_{1}\) & \(J_2\) & \(J_3\) & \(J_4\) & \(t_{K}\) \\
\hline
Cyl 2D      & 2  & 199  & 449  & -   & 151 \\
Cyl 3D      & 3  & 100  & 40   & 64  & 299 \\
Experimental cyl         & 2  & 111    & 301 & -  & 4000 \\

\hline
\end{tabular}
\end{table}

\begin{table}[h!]
\centering
\caption{Dataset splitting for training, validation, and test sets.}
\begin{tabular}{lccc}
\hline
\textbf{Database} & \textbf{Training } & \textbf{Validation } & \textbf{Test } \\
\hline
Cyl 2D            & 106                           & 23                                & 22                          \\
Cyl 3D            & 209                           & 44                                & 45                          \\
Experimental cyl  & 2800                          & 600                               & 600                         \\
\hline
\end{tabular}
\label{tab:trainvalidationtest}
\end{table}

\section{Results and discussion}\label{results and discussion}

This section presents the results obtained from applying the proposed HOSVD-SR methodology to each of the test cases described in Section \ref{secTest}. Various compression rates were considered for each test case to evaluate the method's resolution enhancement capability. The results were compared to those obtained using the SVD-SR approach. Aside from the results comparison of the SVD-SR and HOSVD-SR approaches for different compression rates, the influence of the number of retained modes on the accuracy of the reconstruction has also been compared. This mode truncation has been performed across all available spatial dimensions. The criteria for selecting the number of retained modes depends on the available spatial points in the low-resolution dataset. We start with 5 retained modes for each spatial dimension and then gradually increase the number of modes in increments of five up to the maximum number available. The hyper parameters applied for the training of each neural network can be found in Appendix \ref{annex:anexo}. 

\subsection{Two-dimensional laminar flow past a circular cylinder}

This section presents the results acquired by applying the proposed HOSVD-based resolution enhancement methodology (HOSVD-SR) to the two-dimensional flow past a circular cylinder database obtained by numerical simulation. These results are compared with those achieved using the SVD-SR resolution enhancement method. Additionally, the impact of the number of retained modes on result accuracy was analysed for databases with different compression rates. The compression ratios for each of the databases analysed are presented in Tab. \ref{tab:dimensions_2dcyl}.

\begin{table}[H]
\centering
\caption{Compression ratio and dimensions of the databases under study. \( J_{1} \) refers to the number of variables. \( J_2 \), \( J_3 \), and \( J_4 \) represent the number of points in the \(x\), \(y\), and \(z\) space dimensions, and \( t_K \) represents the number of available snapshots.}
\label{tab:dimensions_2dcyl}
\begin{tabular}{lcccccc}

\textbf{ID} & \textbf{compression ratio } & \(J_{1}\) & \(J_2\) & \(J_3\) & \(J_4\) & \(T_{K}\) \\
\hline
CR1 & 1:1        & 2  & 199    & 449 & -  & 151 \\
CR10 & 1:10        & 2  & 20    & 45 & -  & 151 \\
CR20 & 1:20      & 2  & 10  & 23  & -   & 151 \\
CR40 & 1:40      & 2  & 5  & 12   & -  & 151 \\
CR80 & 1:80      & 2  & 3  & 6   & -  & 151 \\

\hline
\end{tabular}
\end{table}

Table \ref{tab:cyl2d_rmse} presents the reconstruction RRMSE for the low resolution databases detailed in Tab. \ref{tab:dimensions_2dcyl}, using both the SVD-based and HOSVD-based approaches. In all tests, the proposed HOSVD-based methodology consistently yields lower RRMSE values compared to the SVD-SR approach, for both the streamwise and normal velocity components. This improvement is attributed to the HOSVD's ability to independently treat each dimension of the tensor within factor matrices, as opposed to the SVD approach, which combines multiple dimensions into one for decomposition.

For both approaches, the RRMSE for the normal velocity component is consistently higher than for the streamwise component. This discrepancy is likely due to the abrupt changes between the magnitudes of the streamwise and normal velocity components, which complicate the accurate reconstruction of the velocity field. The larger magnitudes of the streamwise component result in higher singular values, leading the method to prioritize its reconstruction. In contrast, the normal velocity component, having lower magnitudes and values close to zero, contributes less to the dominant singular values, making its reconstruction more challenging. As the RRMSE is a relative error measure, the smaller magnitudes of the normal velocity component amplify the discrepancy, resulting in higher relative errors for this component. Moreover, since the normal component exhibits values close to zero throughout most of the domain, even small absolute deviations can produce disproportionately large relative errors, further increasing the reconstruction RRMSE.
Additionally, it is observed that as the compression ratio increases, the RRMSE for both velocity components also increases. This trend is likely a result of information loss during the compression process. Specifically, for the dataset with a compression ratio of 1:40, the HOSVD-based method reconstruction RRMSE is approximately 1.08 and 1.47 times lower for the streamwise and normal velocity components, respectively, compared to the SVD-SR approach. As the compression ratio decreases, the accuracy improvement offered by the HOSVD-based method becomes even more pronounced, with reconstruction RRMSE 1.6 times lower for the streamwise velocity component and 1.45 times for the normal velocity component compared to SVD-SR.  For compression ratios above 1:40, the reconstruction RRMSE increases for both SVD-SR and HOVSD-SR approaches approximately 8 times, this is related to the few amount of SVD modes available in the low-resolution database (which are 3 and 6 for the $x$ and $y$ dimensions) which do not contain enough information to reconstruct the database properly.

\begin{figure}[H]
    \centering
    \includegraphics[width=\textwidth]{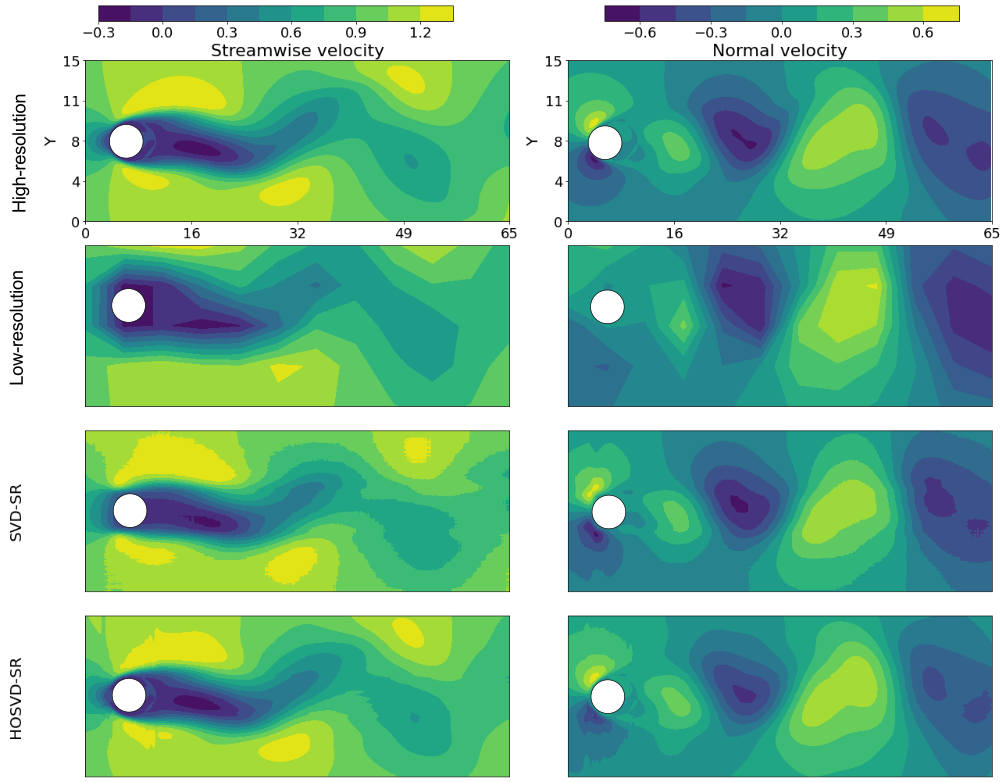} 
    \caption{Two-dimensional flow past a circular cylinder (Re = 100), CR40: Comparison between the SVD-SR and the HOSVD-SR approaches without the truncation of the spatial mode matrices.}
    \label{fig:cyl2d_1_40}
\end{figure}

Figure \ref{fig:cyl2d_1_40} shows the reconstructions of the 2D laminar flow past a circular cylinder obtained using both the SVD-SR and the novel HOSVD-based approach, with an input database with compression ratio of 1:40, with 5 and 12 SVD modes in the $x$ and $y$ dimensions respectively. Both methodologies successfully reconstruct the database with an RRMSE of less than 5\% for both the streamwise and normal velocity components. The reconstructions obtained with the SVD-SR show a mismatch in the flow at the vicinity of the bluff cylinder and inside the von Kármán, while the reconstructions obtained with the HOSVD-SR novel approach are capable of accurately recreating the flow near the cylinder and inside the wake.

\begin{table}[H]
    \centering
    \small 
    \renewcommand{\arraystretch}{2} 
    \caption{Reconstruction RRMSE obtained at different compression rates for the two-dimensional flow past a circular cylinder at Re = 100 using both the HOSVD-SR and SVD-SR approaches. "Variable 1" stands for the "streamwise" velocity component, while "Variable 2" refers to the "normal" one.}
    \label{tab:cyl2d_rmse}
    \begin{tabular}{@{}>{\centering\arraybackslash}m{1.5cm} >{\centering\arraybackslash}m{3cm}| 
    >{\centering\arraybackslash}m{2.5cm} >{\centering\arraybackslash}m{2.5cm}| 
    >{\centering\arraybackslash}m{2.5cm} >{\centering\arraybackslash}m{2.5cm}@{}}
        \hline
        \multicolumn{2}{c|}{Compression rate} & \multicolumn{2}{c|}{SVD-SR \quad ($\times 10^{-2}$)} & \multicolumn{2}{c}{HOSVD-SR \quad ($\times 10^{-2}$)} \\ 
        \multicolumn{2}{c|}{} & RRMSE variable 1 & RRMSE variable 2 & RRMSE variable 1 & RRMSE variable 2  \\ \hline
        CR10 & \vspace{0.5cm}\includegraphics[width=3cm]{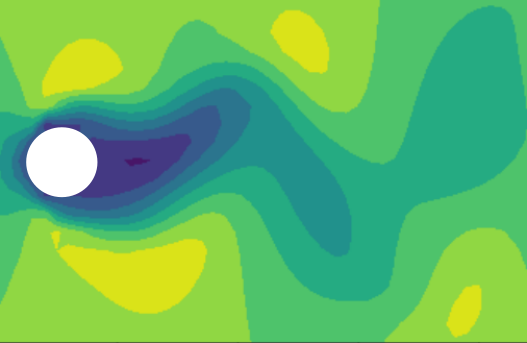}\vspace{0.5cm} & 1.18 & 1.28 &  0.45 & 0.56    \\ \hline
        CR20 & \vspace{0.5cm}\includegraphics[width=3cm]{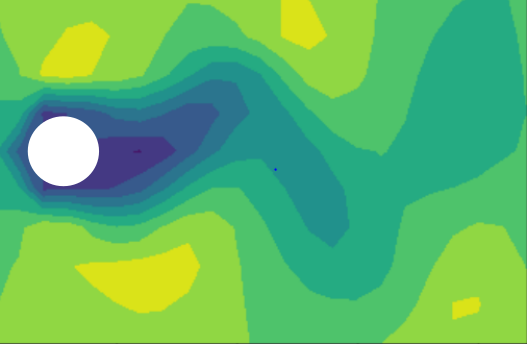}\vspace{0.5cm} & 1.37 & 1.86 &  0.94 & 1.13 \\ \hline
        CR40 & \vspace{0.5cm}\includegraphics[width=3cm]{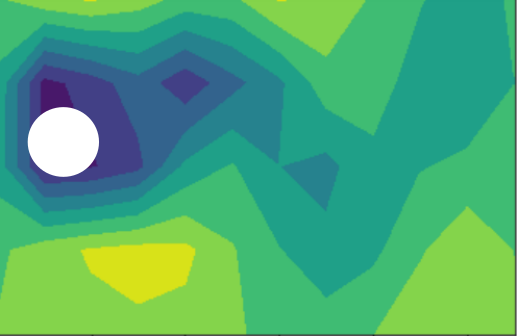}\vspace{0.5cm} & 1.45 & 2.84 & 1.34 & 1.93 \\ \hline
        CR80 & \vspace{0.5cm}\includegraphics[width=3cm]{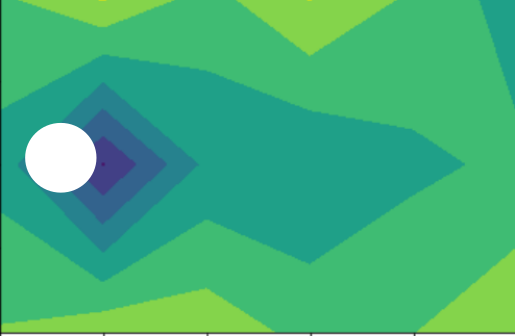}\vspace{0.5cm} & 7.22 & 11.4 & 6.37 & 8.06 \\ \hline
    \end{tabular}
\end{table}

\begin{figure}[H]
    \centering
    \includegraphics[width=\textwidth]{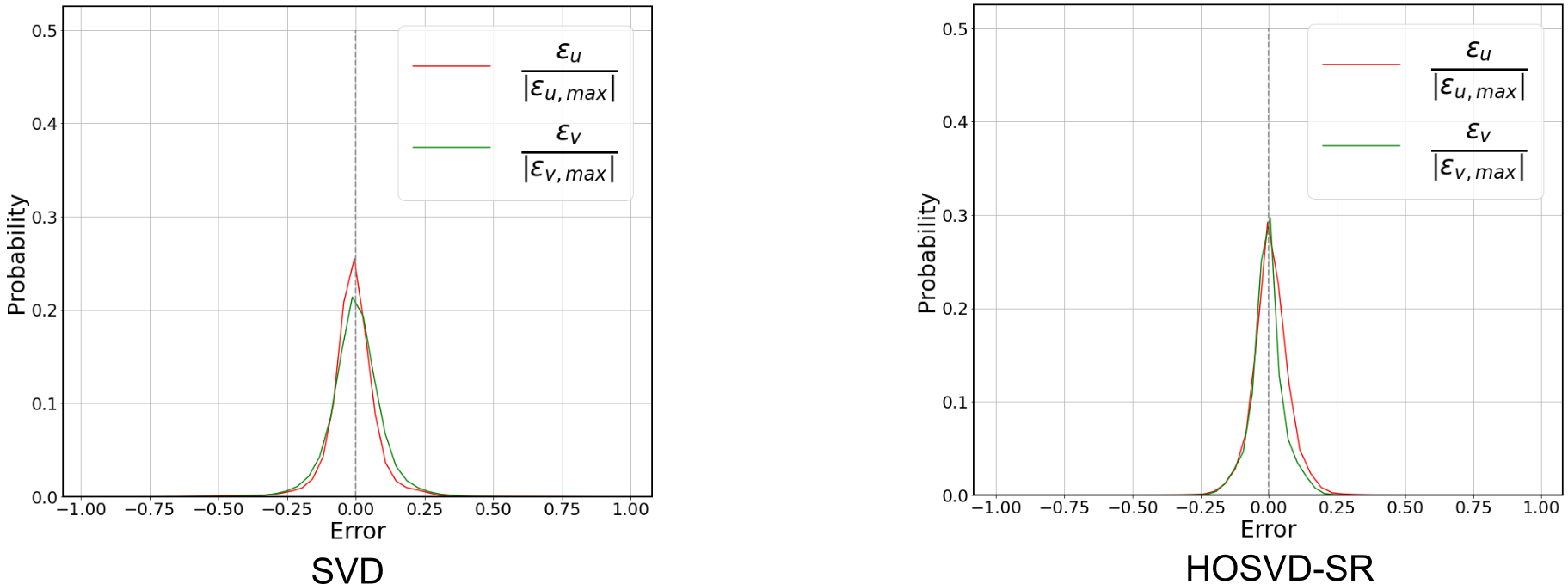} 
    \caption{Two-dimensional flow past a circular cylinder (Re = 100), CR40: normalized error estimated probability distribution function of both the SVD-SR and the HOSVD-SR approaches.}
    \label{fig:cyl2d_1_40_pdf}
\end{figure}

Figure \ref{fig:cyl2d_1_40_pdf} presents the normalized error probability distribution function for the streamwise and normal velocity components in the reconstruction of the 1:40 compression ratio dataset, using both the SVD-SR and HOSVD-SR approaches. The obtained probability distribution functions of the normalized error for the SVD-SR approach exhibit a broad normal distribution centred around 0. For both the normal and streamwise velocity components, the probability of the normalized error being exactly 0 is approximately 20\%. The wide spread and relatively low peak suggest a greater dispersion in the error distribution, indicating less around 0. In contrast, the HOSVD-SR probability distribution functions also exhibit a narrower distribution for both normal and streamwise velocity components, with peaks around 0, suggesting a smaller dispersion of the error in the reconstruction. The probabilities of the normalized error being 0 for the normal and streamwise velocity components are 30\% for the normal velocity components respectively.

\begin{figure}[H]
    \centering
    \includegraphics[width=\textwidth]{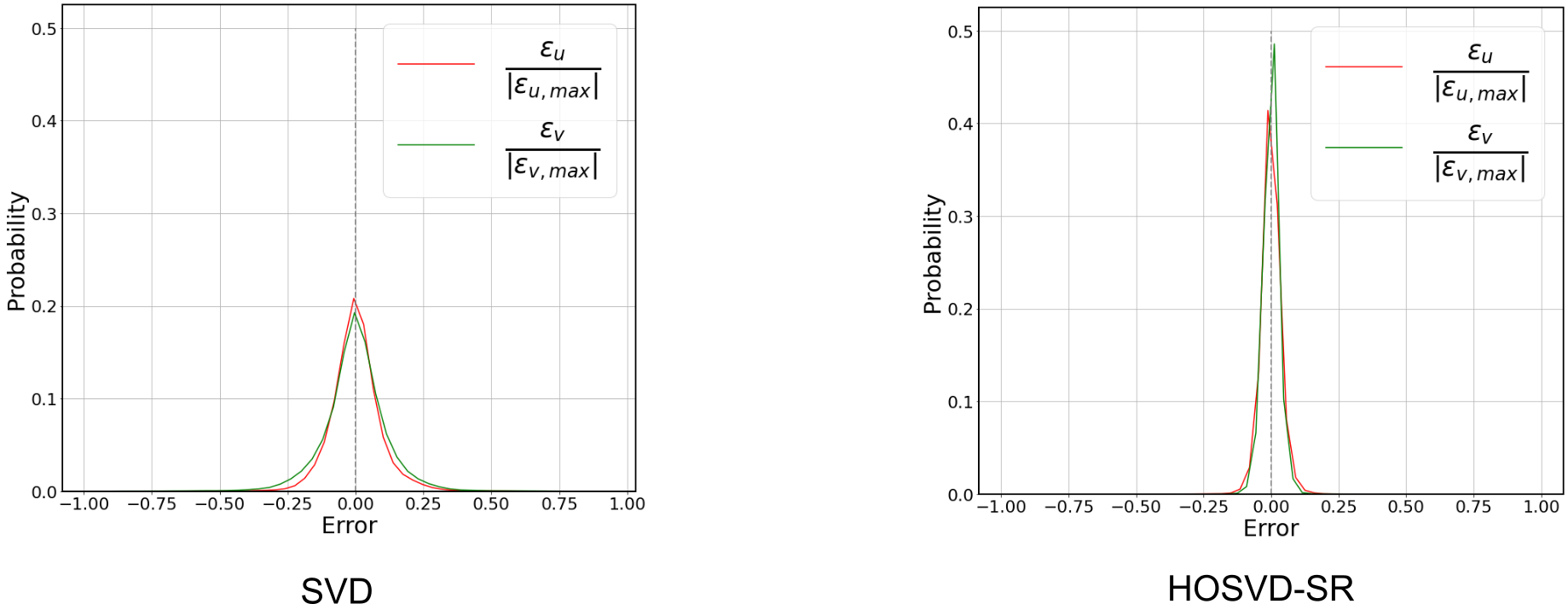} 
    \caption{
    Counterpart of Fig. \ref{fig:cyl2d_1_40_pdf} but for CR20.}
    \label{fig:cyl2d_1_20_pdf}
\end{figure}

The probability distribution functions of the reconstructions obtained using the HOSVD-SR and the SVD-SR approaches for the 1:20 low resolution database is shown in Fig. \ref{fig:cyl2d_1_20_pdf}. The SVD-based approach exhibits a normal distribution with a high standard deviation, with approximately a 20\% probability for the normalized error to be zero for both normal and streamwise velocity components. In contrast, the PDF for both the streamwise and normal velocity components using the proposed HOSVD-SR approach is represented by a narrow normal distribution, with probabilities for the normalized error to be equal to zero of 41\% and 48\%, respectively. Thus, the proposed approach outperforms the SVD-based one, achieving lower standard deviations for the velocity component reconstructions. 

The probability distribution functions of the reconstruction obtained using the SVD-SR approach for the database with compression ratio 1:20 shows lower peaks compared to the one with a compression ratio of 1:40, which as explained by \cite{diaz_morales2024} is related to overfitting during the training of the neural network due to the large amount of data. As seen, using HOSVD-SR overcomes the problem of overfitting, presenting a more robust and generalizable model.

For the two-dimensional flow past a circular cylinder (Re $=$ $100$) several tests were conducted to determine the minimum number of SVD modes required for accurately reconstructing the original high-resolution database using HOSVD-SR. Given that the number of retained SVD modes is limited by the number of points in each spatial dimension, the 1:20 compression ratio database was selected to illustrate the influence of retained SVD modes and assess the methodology's compression capabilities. For lower compression ratios, there are more spatial points available in the data base which means that it is possible to select a higher number of SVD modes, and also perform any truncation if it is required. 

Table \ref{tab:2dcyl_modes} presents the reconstruction RRMSE values, calculated between the high resolution tensor and the reconstructions obtained through both SVD and HOSVD-SR approaches, for all the tests performed with different compression ratios and retained SVD modes. It is worth clarifying that the maximum number of retained SVD modes on each spatial dimension is equal to its number of points, making it feasible to reconstruct the resolution enhanced database with fewer SVD modes.The reconstruction RRMSE values  shown in Tab. \ref{tab:2dcyl_modes} demonstrate a direct correlation between the compression ratio and the accuracy of the reconstruction, retaining a fixed number of SVD modes on each spatial dimension ($5$, $10$, $15$ and $20$ SVD modes). The reconstruction RRMSE is lower as long as the compression ratio decreases, due to the fact that the input contains more information and is closer to the original dataset, which enables the neural network to provide precise weights for the reconstruction. Regarding the amount of retained SVD modes for each compression ratio, the RRMSE decreases as long as more SVD modes are taken for the reconstruction. As mentioned before, in some cases retaining the maximum number of available SVD modes can lead to over fitting in the network. Diaz, et al. ~\cite{diaz_morales2024} in their work demonstrated the effect that the number of SVD modes retained has in the accuracy of the reconstruction. Thus, a sensibility analysis might be performed over the database to choose the adequate number of retained SVD modes for an accurate reconstruction. 

\begin{table}[H]
\caption{Two-dimensional flow past a circular cylinder (Re = 100), HOSVD-SR reconstruction RRMSE for the streamwise velocity component with various spatial SVD modes retained.}
\centering
\begin{tabular}{lcccc}
\toprule
\multicolumn{5}{c}{\textbf{RRMSE} $\times 10^{-2}$} \\
\midrule
\multirow{2}{*}{\textbf{compression ratio}} & \multicolumn{4}{c}{\textbf{Number of retained SVD modes}} \\
          & 5 & 10 & 15 & 20 \\
\midrule
1:10      & 5.03 & 1.97 & 1.34 & 0.45 \\
1:20      & 6.11 & 0.94 & - & - \\
1:40      & 1.34 & - & - & - \\
\bottomrule
\end{tabular}
\label{tab:2dcyl_modes}
\end{table}

\begin{figure}[H]
    \centering
    \includegraphics[width=\textwidth]{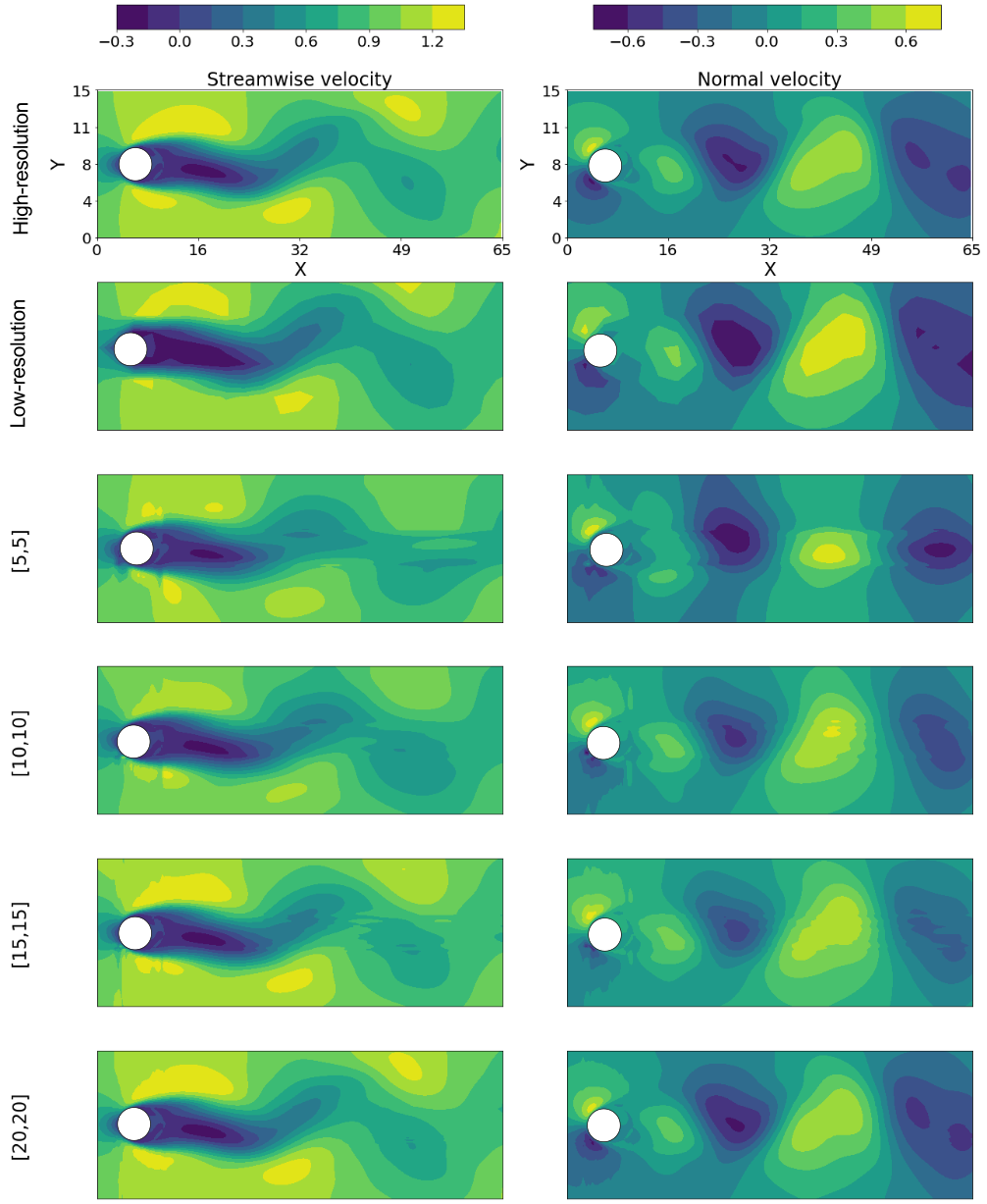} 
    \caption{Two-dimensional flow past a circular cylinder (Re = 100), CR20: Comparison between the HOSVD-SR reconstructions  with: $5$, $10$, $15$ and $20$ spatial SVD modes retained on each spatial dimension.}
    \label{fig:cyl2d_modes}
\end{figure}

Figure \ref{fig:cyl2d_modes} shows the reconstructions of the streamwise and normal velocity components with $5$, $10$, $15$, and $20$ retained SVD modes on each one of the spatial coordinates. The reconstructions obtained by retaining 15 SVD modes in each spatial dimension show an RRMSE of $1.46$ $\times$ \(10^{-2}\) in the streamwise velocity component. As seen,  the cylinder wake  flow structures within the von Kármán street and near the solid body for both velocity components were accurately reconstructed.

With $10$ SVD spatial modes retained, the reconstruction successfully captures the nature of the phenomenon, achieving an RRMSE of $1.97$ $\times$ \(10^{-2}\) in the streamwise velocity component. However, the flow structures within the von Kármán street appear slightly merged, possibly due to truncation performed on the mode matrices in the spatial dimensions. This trend is further evident in the results obtained when retaining only $5$ spatial SVD modes in the $x$ and $y$ dimensions, where the flow structures within the wake are more significantly merged, and the flow in the vicinity of the bluff cylinder are not accurately reconstructed.  

\subsection{Three-dimensional flow past a circular cylinder}

This section exhibits the results and analysis of the application of the HOSVD-SR resolution enhancement methodology to the low-resolution datasets generated from the three-dimensional flow past a circular cylinder database. The analysis relies on the calculation of the reconstruction RRMSE for each compression ratio tested (detailed in Tab. \ref{tab:dimensions_3dcyl}), and its comparison with the results obtained using the SVD-SR approach. The effects of retaining different numbers of spatial SVD modes on the reconstruction accuracy has been studied. 

\begin{table}[h!]
\centering
\caption{ Same as in Tab. \ref{tab:dimensions_2dcyl} but for the three-dimensional flow past a circular cylinder.}
\label{tab:dimensions_3dcyl}
\begin{tabular}{lcccccc}

\textbf{ID} &\textbf{compression ratio } & \(J_1\) & \(J_2\) & \(J_3\) & \(J_4\) & \(t_K\) \\
\hline
CR1 & 1:1        & 3  & 100    & 40 & 64  & 299 \\
CR2& 1:2        & 3  & 50    & 20 & 32  & 299 \\
CR4 & 1:4      & 3  & 25  & 10  & 16   & 299 \\
CR20 & 1:20      & 3  & 5  & 2   & 3  & 299 \\

\hline
\end{tabular}
\end{table}

The reconstruction RRMSE for various compression ratios applied to the 3D circular cylinder database using both the SVD-SR approach and the novel HOSVD-based method proposed is detailed in Tab. \ref{tab:cyl3d_rmse}. It can be observed that the HOSVD-SR approach yields greater accuracy for each variable analysed (streamwise, normal, and spanwise velocity components) compared to its counterpart. Both approaches show a decrease in RRMSE as the compression ratio decreases. Notably, the RRMSE values corresponding to the third variable, obtained with the 1:2 and 1:4 compression ratios, are very close to each other (around 1$\times10\textsuperscript{-3}$), indicating the best accuracy achieved for this specific variable.

For the database with a compression ratios of 1:20, with 5, 2 and 3 spatial points in the $x$, $y$ and $z$ dimensions respectively, the HOSVD-SR method is able to provide reconstructions with RRMSE values $1.41$ times lower on the streamwise velocity component and $3.10$ times lower normal velocity component, than the SVD-SR approach. The 1:10 compression ratio database shows close-to-equal RRMSE values in both normal and spanwise velocity components using both SVD-SR and HOSVD-SR methods, while for the streamwise velocity component the RRMSE is approximately $1.78$ times lower using HOSVD-SR.  
\begin{table}[H]
    \centering
    \small 
    \renewcommand{\arraystretch}{2} 
    \caption{ Same as Tab. \ref{tab:cyl2d_rmse} but for the three-dimensional flow past a circular cylinder at Re = 280.}
    \label{tab:cyl3d_rmse}
    \begin{tabular}{@{}>{\centering\arraybackslash}m{1cm} >{\centering\arraybackslash}m{3cm}| 
    >{\centering\arraybackslash}m{1.5cm} >{\centering\arraybackslash}m{1.5cm} >{\centering\arraybackslash}m{1.5cm}| 
    >{\centering\arraybackslash}m{1.5cm} >{\centering\arraybackslash}m{1.5cm} >{\centering\arraybackslash}m{1.5cm} 
    >{\centering\arraybackslash}m{1.5cm} >{\centering\arraybackslash}m{1.5cm}@{}}
        \hline
        \multicolumn{2}{c|}{Compression rate} & \multicolumn{3}{c|}{SVD-SR \quad ($\times 10^{-2}$)} & \multicolumn{3}{c}{HOSVD-SR \quad ($\times 10^{-2}$)} \\ 
        \multicolumn{2}{c|}{} & RRMSE v. 1 & RRMSE v. 2 & RRMSE v. 3 & RRMSE v. 1 & RRMSE v. 2 & RRMSE v. 3  \\ \hline
        CR2 & \vspace{0.5em}\includegraphics[width=3cm]{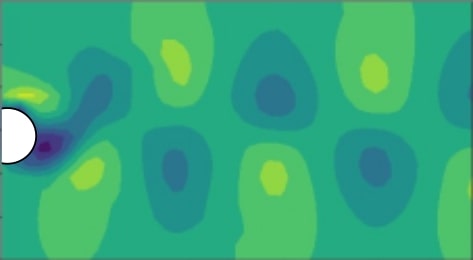}\vspace{0.5em} & 6.63 & 8.60 & 7.72 & 4.42 & 8.34 & 7.72 \\ \hline
        CR4 & \vspace{0.5em}\includegraphics[width=3cm]{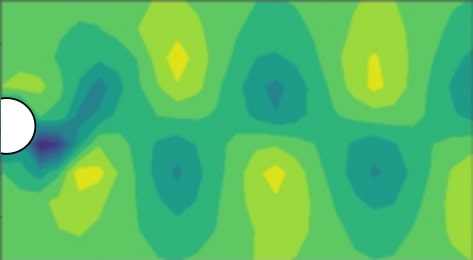}\vspace{0.5em} & 7.98 & 8.02 & 7.72 & 4.48 & 8.41 & 7.82 \\ \hline
        CR20 & \vspace{0.5em}\includegraphics[width=3cm]{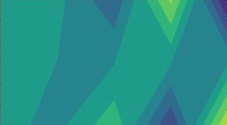}\vspace{0.5em} & 10.14 & 26.82 & 8.31 & 7.19 & 8.63 & 8.48 \\ \hline
    \end{tabular}
\end{table}

\begin{figure}[H]
    \centering
    \includegraphics[width=\textwidth]{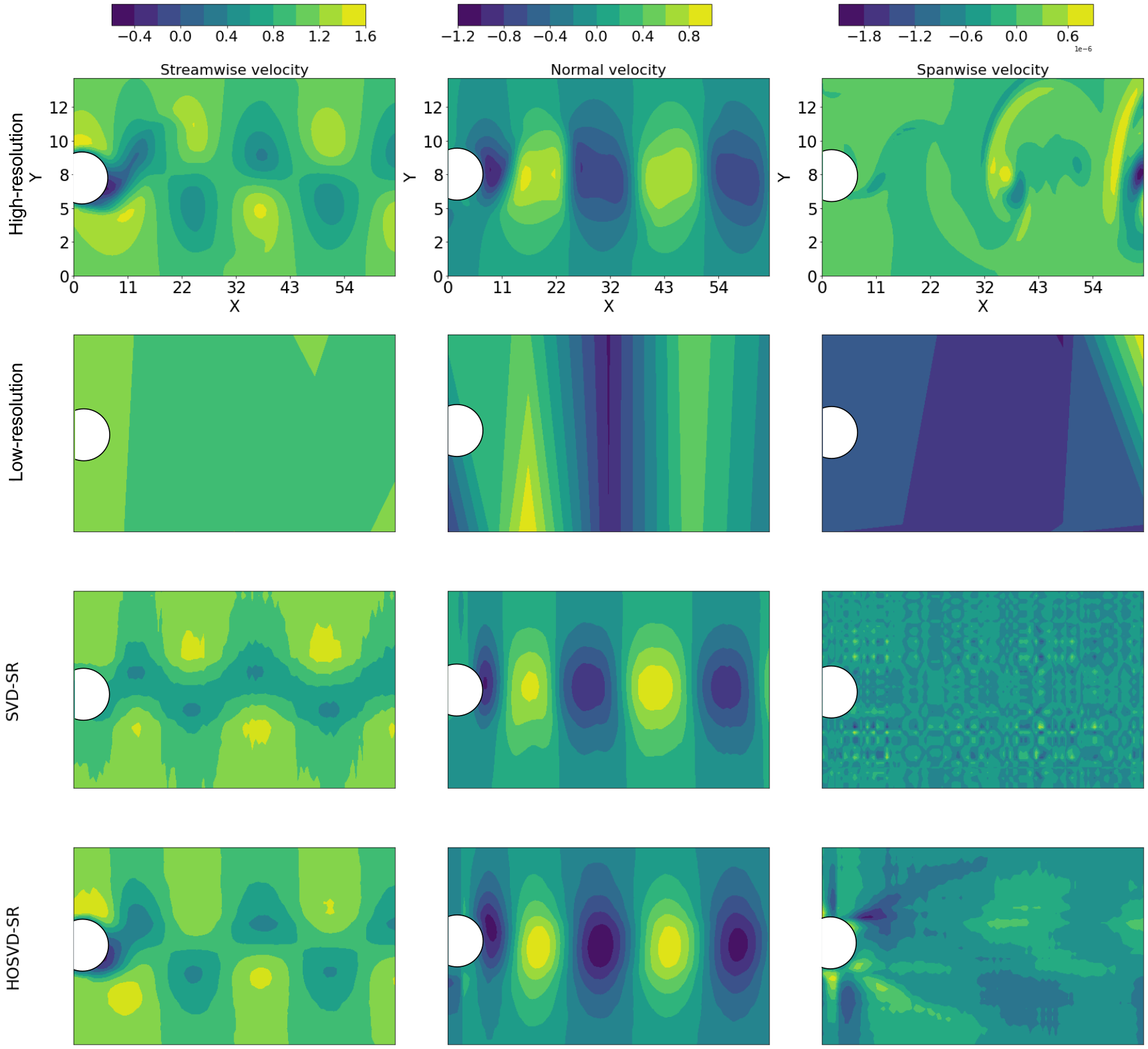} 
    \caption{Counterpart of Fig. \ref{fig:cyl2d_1_40} but for the three-dimensional flow past a circular cylinder (Re = 280), CR20.}
    \label{fig:cyl3d_5}
\end{figure}

Figure \ref{fig:cyl3d_5} presents the reconstructions obtained from an input dataset with a compression ratio of 1:20 using both the SVD-SR approach and the HOSVD-SR method. The SVD-based method successfully recreates the general von Kármán vortex street but struggles with small-scale structures. It fails to accurately reconstruct the bluff body's surrounding flow structures in the streamwise velocity component. In contrast, HOSVD-SR accurately captures the phenomenon using only $10$ SVD modes in the streamwise, normal, and spanwise directions. However, despite their low RRMSE values, neither the SVD-SR nor the HOSVD-SR are able to reconstruct the spanwise velocity component accurately. This limitation is likely due to the small magnitude of the spanwise velocity component (on the order of $1 \times 10^{-6}$) compared with the values encountered in the other components. 

To further investigate this issue, the dataset was standardized by centering and scaling each velocity component to zero mean and unit variance. This normalization gives equal weight to each variable during decomposition, which is particularly beneficial when handling multiple variables of different orders of magnitude. This strategy has proven effective in similar contexts, as reported by Hetherington et al. in Ref.~\cite{HETHERINGTON2024109217}. However, the spanwise velocity is under development in the dataset used for the training of the model. In order to reconstruct this component, a database where the spanwise velocity component is more developed should be used, but this is far from the scope of the current work and will remains open for future research.

\begin{figure}[H]
    \centering
    \includegraphics[width=\textwidth]{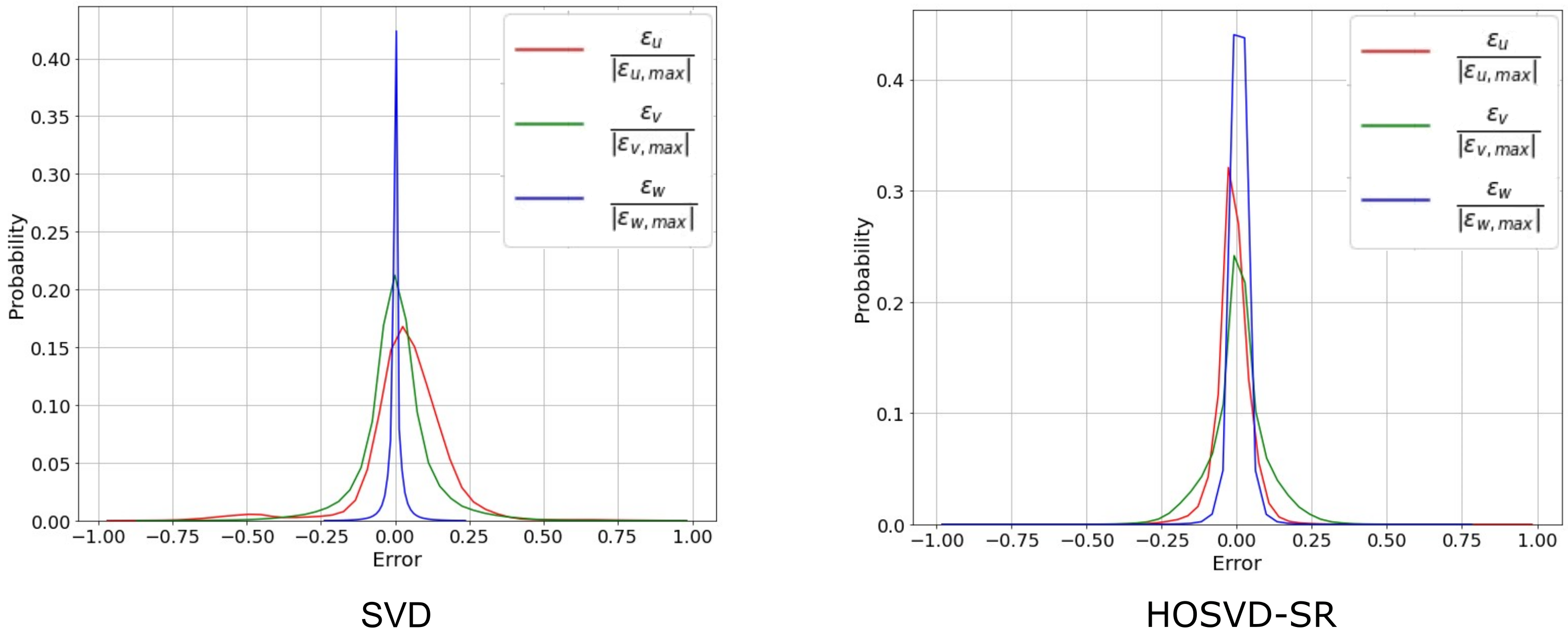} 
    \caption{Same as Fig. \ref{fig:cyl2d_1_40_pdf} but for the three-dimensional flow past a circular cylinder (Re = 280),  CR20.}
    \label{fig:cyl3d_5_pdf}
\end{figure}

Figure \ref{fig:cyl3d_5_pdf} compares the probability distribution functions obtained by calculating the normalized error for the three variables under study: streamwise, normal, and spanwise velocity for the input tensor with a compression ratio of 1:20. The SVD-SR reconstruction shows lean normal distributions centred around $0$ for both the normal and spanwise velocity components, with maximum occurrence probabilities of approximately $42\%$ and $20\%$, respectively. For the spanwise velocity component, this distribution is related to the scale of its magnitudes. The streamwise velocity component is described by a right-skewed distribution with a small peak located at $5$ $\times$ $10^{-1}$ along the error axis, with a maximum probability of occurrence around $16\%$ for errors around 2 $\times$  $10^{-3}$.

On the other hand, the HOSVD-SR reconstruction probability distribution functions for the streamwise and normal velocity components correspond to an asymmetric right-skewed distribution. The maximum probability of occurrence is $32\%$ at an error of $2.5$ $\times$  $10^{-2}$  for the streamwise velocity component, and a peak of $24\%$ at an approximate error of 5 $\times$  $10^{-3}$  for the normal velocity component. The spanwise velocity component exhibits a left-skewed distribution, with approximately a $42\%$ chance of occurrence for errors around 4 $\times$  $10^{-3}$.

\begin{figure}[H]
    \centering
    \includegraphics[width=\textwidth]{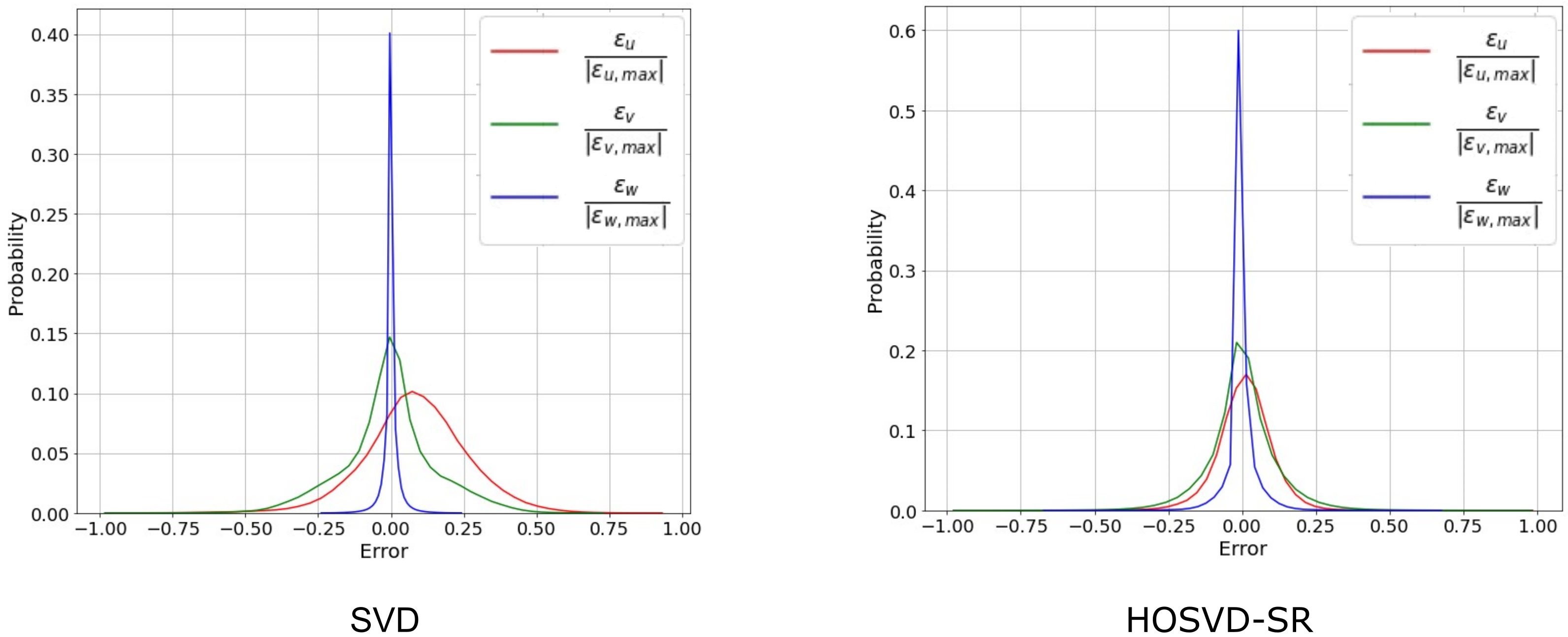} 
    \caption{Counterpart of Fig. \ref{fig:cyl2d_1_40_pdf} but for the three-dimensional flow past a circular cylinder (Re = 280),  CR10.}
    \label{fig:cyl3d_25_pdf}
\end{figure}

Figure \ref{fig:cyl3d_25_pdf} displays the estimated probability density curves of the normalized error for both SVD-SR and HOSVD-SR approaches. The SVD-SR method tends to produce a right-skewed distribution for the streamwise velocity with a 10\% probability for the normalized error to be equal to 8 $\times$ $10^{-2}$. In contrast, the HOSVD-SR method produces a quasi-symmetric distribution centred almost around 0 for both the streamwise and normal velocities. For the spanwise velocity, the PDFs show that HOSVD-SR has a tendency towards small negative errors, while the SVD-SR errors are more uniformly distributed.

As with the 2D flow past a circular cylinder database, the 3D database was obtained through a numerical simulation. Consequently, the number of SVD modes retained is limited to the spatial data of the input tensor; therefore, the input database must have a sufficient number of spatial points to perform this test. Among the various compression rates applied to the original database, the one with a ratio of 1:4 was selected to evaluate the data compression capabilities of the method, due to its relatively high number of available SVD modes.

\begin{figure}[H]
    \centering
    \includegraphics[width=\textwidth]{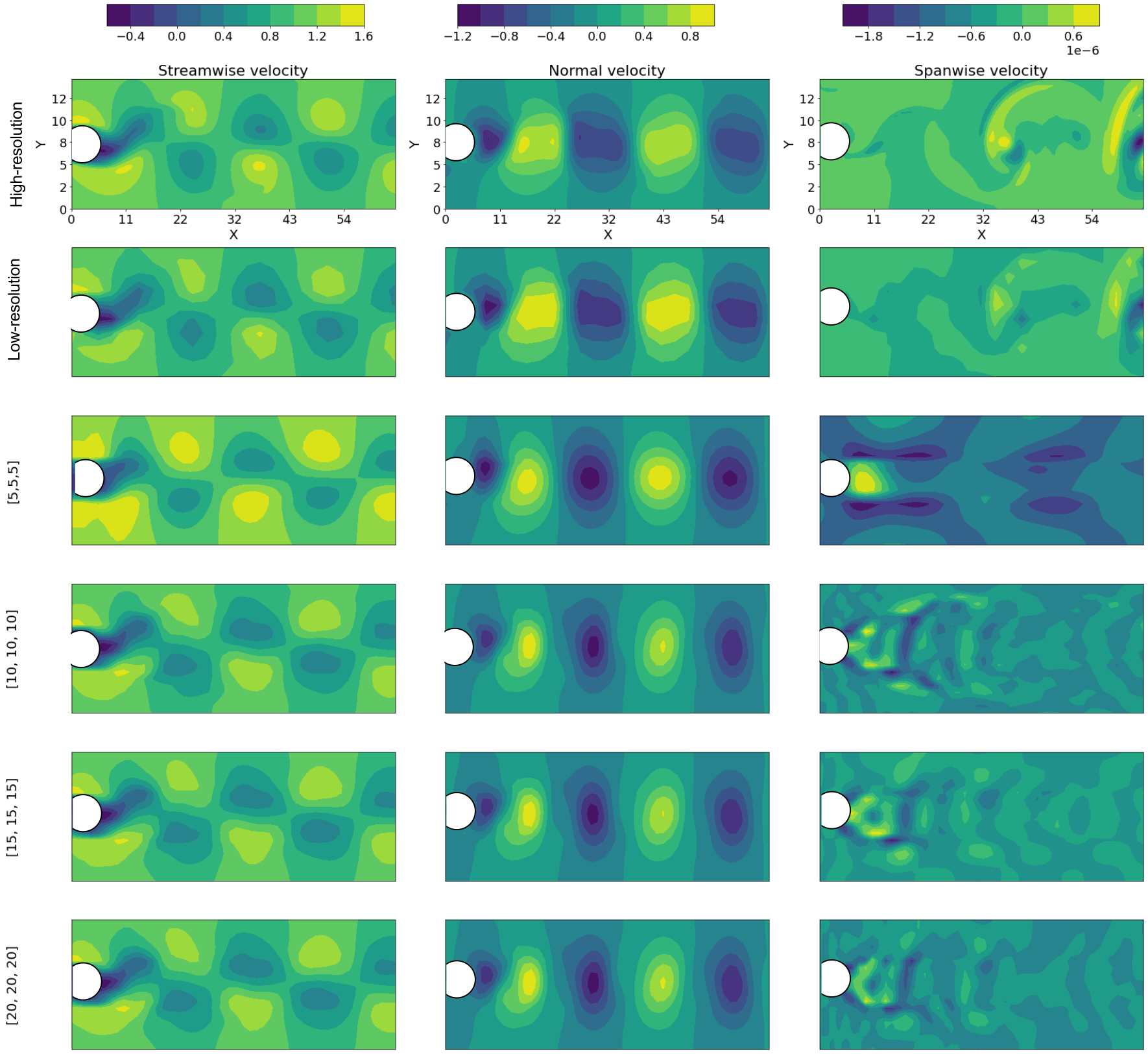} 
    \caption{Three-dimensional flow past a circular cylinder (Re = 280), CR4: Comparison between the HOSVD-SR reconstructions with: 5, 10, 15 and 20 spatial SVD modes retained on each spatial dimension.}
    \label{fig:cyl3d_modes}
\end{figure}

Figure \ref{fig:cyl3d_modes} displays the reconstructions obtained retaining 5, 10, 15 and 20 SVD modes for each of the $x$, $y$ and $z$ spatial dimensions. It can be observed that for all the streamwise velocity components, the edges of the flow structures are under defined while for the normal velocity component the methodology is capable to reconstruct all the flow structures. It must be clarified that the methodology is unable to accurately reconstruct the spanwise velocity field, this can be related to the small scale of the magnitudes involved in the phenomenon. 

The reconstruction obtained with 5 retained SVD modes in the $x$ spatial dimension is able to recreate the von Kármán phenomenon, though it shows a mismatch between the flow structures within the vortex street and those around the circular cylinder. Specifically, the flow inside the vortex for the streamwise velocity component appear almost regular in shape. Reconstructions using 10 retained spatial SVD modes in the $x$, $y$ and $z$ dimensions also exhibit a mismatch between the flow structures inside the vortex street and those around the solid body. However, the reconstruction obtained with $15$ and $20$ retained SVD modes successfully reproduces the flow structures both near the solid body and within the vortex street. Despite the low difference in the RRMSE of these reconstructions, it is evident that selecting an appropriate number of retained SVD modes can significantly improve the reconstruction accuracy while also compressing the information.

Table \ref{tab:cyl3d_rmse_modes} presents the reconstruction RRMSE values obtained by retaining different numbers of spatial SVD modes. The lowest RRMSE is achieved with a 1:4 compression ratio and $15$ retained SVD modes in the $x$ spatial dimension. Retaining either more or fewer SVD modes leads to slightly higher RRMSE values, indicating that selecting an appropriate number of SVD modes is crucial for optimal reconstruction accuracy. Consistent with the previously described tests, the RRMSE generally decreases as the compression ratio decreases.

\begin{table}[H]
\centering
\caption{ Same as Tab. \ref{tab:2dcyl_modes} but for the three-dimensional flow past a circular cylinder (Re = 280) databases.}
\begin{tabular}{lcccc}
\toprule
\multicolumn{5}{c}{\textbf{RRMSE} $\times 10^{-2}$} \\
\midrule
\multirow{2}{*}{\textbf{compression rate}} & \multicolumn{4}{c}{\textbf{Number of retained SVD modes}} \\
          & 5 & 10 & 15 & 20 \\
\midrule
1:2       & 5.22 & 4.90 & 4.69 & 4.42 \\
1:4       & 5.17 & 4.49 & 4.41 & 4.48 \\
1:20      & 7.18 & - & - & - \\
\bottomrule
\end{tabular}
\label{tab:cyl3d_rmse_modes}
\end{table}

\subsection{Experimental database of turbulent cylinder wake flows}

The results obtained by applying the HOSVD-based resolution enhancement methodology to the experimental database of cylinder wake flows under different compression rates and number of retained SVD modes are described in this section. Two different analyses were performed. Table \ref{tab:svd_vki} shows the reconstruction RRMSE for both SVD and HOSVD approaches for tensors with different compression rates. The proposed HOSVD-based approach consistently achieves RRMSE values approximately half of those obtained using the SVD-SR method across all considered compression rates.

\begin{table}[h!]
\centering
\caption{Counterpart of Tab. \ref{table: database_summary} but for the experimental flow past a circular cylinder. \( J_2 \), \( J_3 \), and \( J_4 \) represent the number of points in the \(x\), \(y\), and \(z\) space dimensions, and \( t_K \) represents the number of available snapshots.}
\label{tab:dimensions}
\begin{tabular}{lcccccc}

\textbf{ID} & \textbf{compression ratio } & \(J_1\) & \(J_2\) & \(J_3\) & \(J_4\) & \(t_K\) \\
\hline
CR1 & 1:1        & 2  & 111    & 301 & -  & 4000 \\
CR2& 1:2        & 2  & 55    & 150 & -  & 4000 \\
CR5 & 1:5      & 2  & 22  & 74  & -   & 4000 \\
CR10 & 1:10      & 2  & 11  & 30   & -  & 4000 \\
CR20 & 1:20    & 2  & 5  & 15   & -  & 4000 \\
\hline
\end{tabular}
\end{table}

\begin{table}[H]
    \centering
    \small 
    \renewcommand{\arraystretch}{2} 
    \caption{Same as Tab. \ref{tab:cyl2d_rmse} but for the experimental database of a circular cylinder wake flow (Re = 2600).}
    \label{tab:svd_vki}
    \begin{tabular}{@{}>{\centering\arraybackslash}m{1.5cm} >{\centering\arraybackslash}m{3cm}| 
    >{\centering\arraybackslash}m{2.5cm} >{\centering\arraybackslash}m{2.5cm}| 
    >{\centering\arraybackslash}m{2.5cm} >{\centering\arraybackslash}m{2.5cm}@{}}
        \hline
        \multicolumn{2}{c|}{Compression rate} & \multicolumn{2}{c|}{SVD-SR \quad ($\times 10^{-2}$)} & \multicolumn{2}{c}{HOSVD-SR \quad ($\times 10^{-2}$)} \\ 
        \multicolumn{2}{c|}{} & RRMSE v. 1 & RRMSE v. 2 & RRMSE v. 1 & RRMSE v. 2  \\ \hline
        CR2 & \vspace{0.5em}\includegraphics[width=3cm]{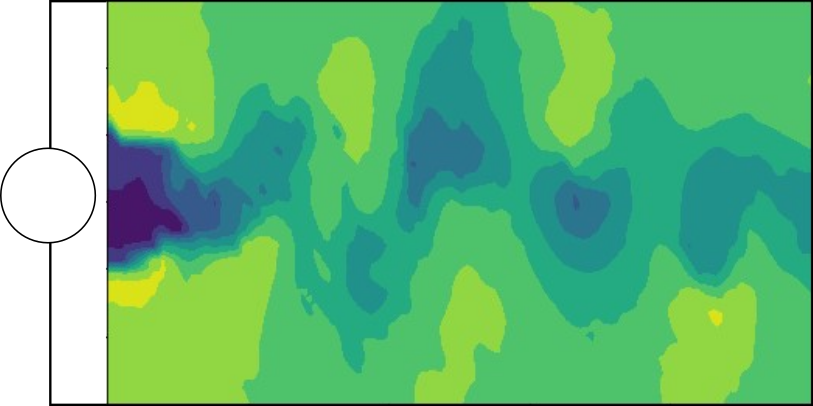}\vspace{0.5em} & 6.81 & 9.05 & 1.17 & 1.74 \\ \hline
        CR5 & \vspace{0.5em}\includegraphics[width=3cm]{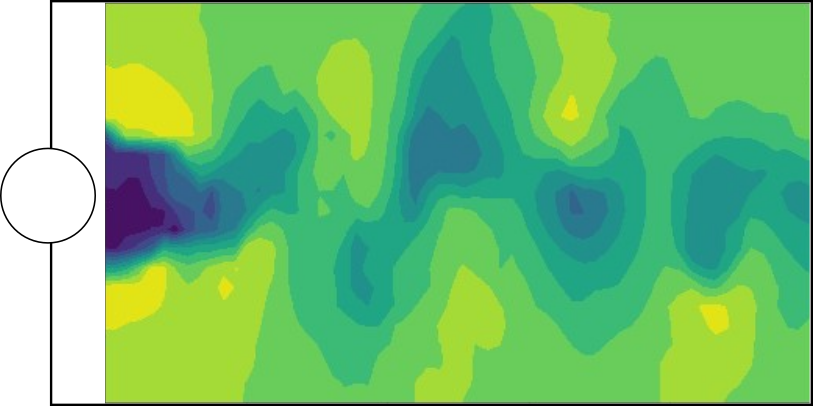}\vspace{0.5em} & 8.31 & 11.76 & 2.93 & 3.35 \\ \hline
        CR10 & \vspace{0.5em}\includegraphics[width=3cm]{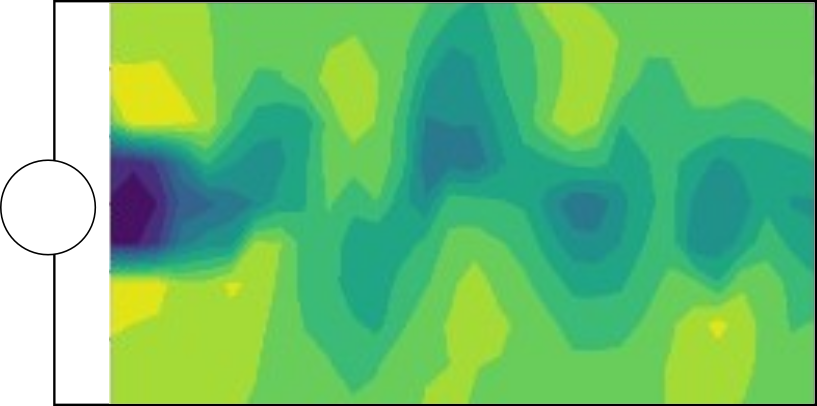}\vspace{0.5em} & 7.54 & 12.73 & 3.21 & 3.81 \\ \hline
        CR20 & \vspace{0.5em}\includegraphics[width=3cm]{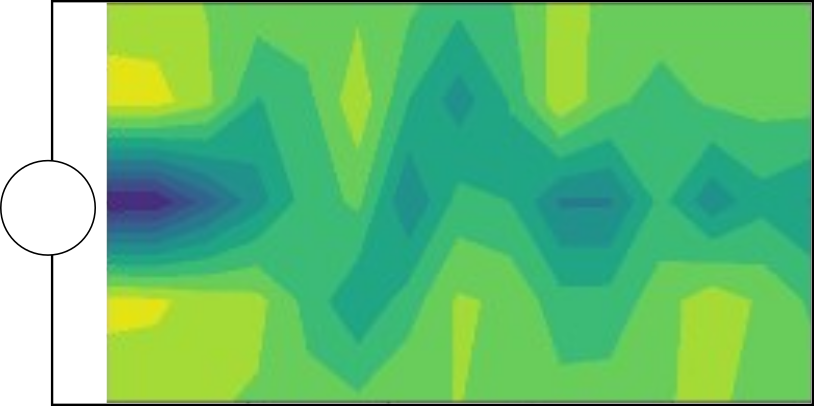}\vspace{0.5em} & 8.18 & 14.03 & 4.55 & 4.79  \\ \hline
    \end{tabular}
\end{table}

It can be observed that the RRMSE for both the streamwise and normal velocity components increases with higher compression rates in both the SVD-SR and the HOSVD-SR proposed methodology. Specifically, the RRMSE for the normal velocity component is almost double that of the streamwise component for both methods. This discrepancy is related to the nature of the problem; as shown in Fig. \ref{fig:vki}, the magnitudes of the normal velocity component exhibit smaller scales (closer to 0) than the ones encountered in the streamwise velocity component, which complicates the reconstruction process due to the data organization of the SVD. 
  
\begin{figure}[H]
    \centering
    \includegraphics[width=14 cm ]{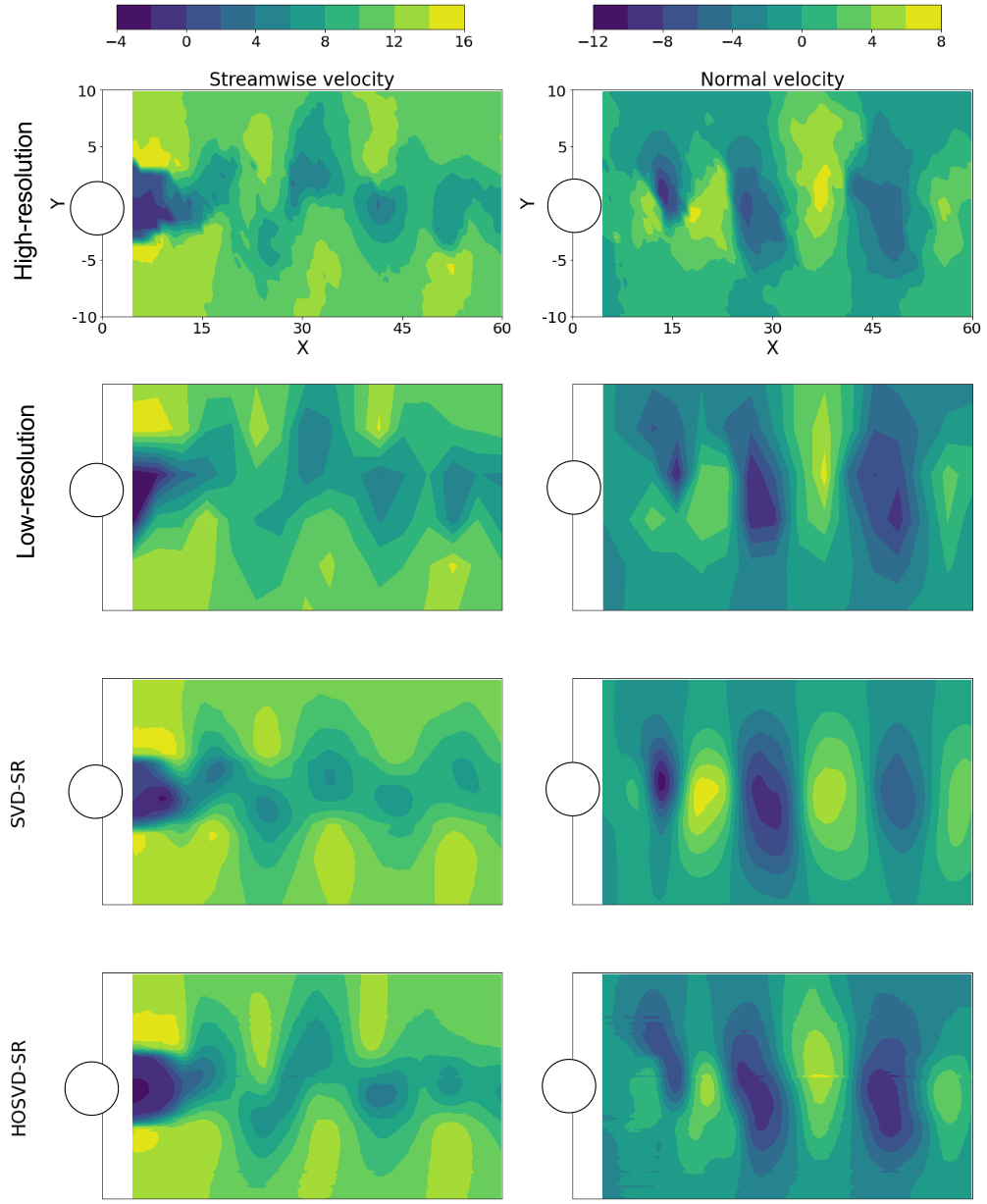} 
    \caption{Same as Fig. \ref{fig:cyl2d_1_40} but for the experimental circular cylinder wake flows (Re = 2600), CR20 database.}
    \label{fig:5_VKI}
\end{figure}

Figure \ref{fig:5_VKI} illustrates the reconstruction results for a compression ratio of 1:20 applied to the experimental database.  The SVD-SR reconstruction RRMSE errors are $8.18$ $\times$ $10^{-2}$ for the streamwise velocity component and $14.3$ $\times$ $10^{-2}$ for the normal velocity component. In comparison, the HOSVD-based approach achieves RRMSE values of $4.55$ $\times$ $10^{-2}$ for the streamwise velocity and $4.79$ $\times$ $10^{-2}$ for the normal velocity component. For compression ratio of 1:20 , both methods smoothly reconstruct the von Kármán wake, displaying the flow structures with regular shapes, demonstrating the de-noising capabilities of HOSVD. The HOSVD-based enhancement method captures the irregularities at the wake's end, whereas the SVD approach yields a quasi-symmetrical shape.

\begin{figure}[H]
    \centering
    \includegraphics[width=\textwidth]{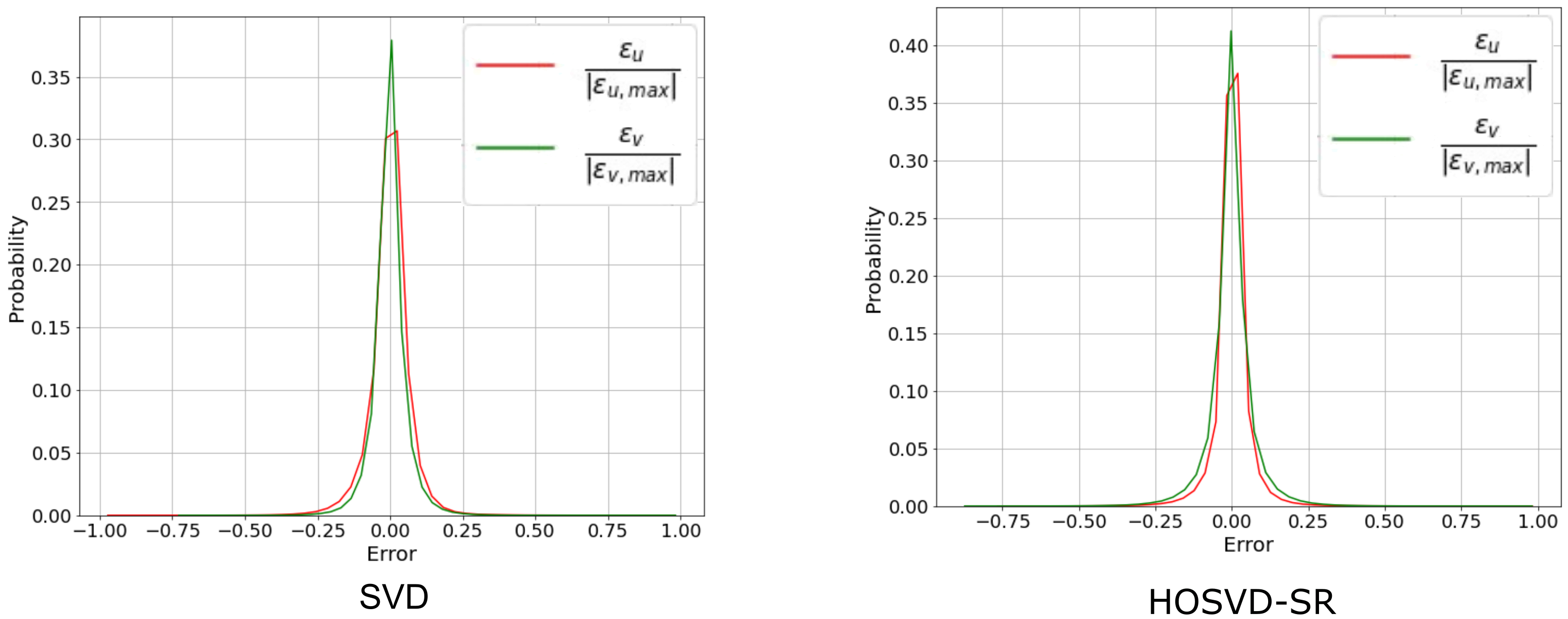} 
    \caption{ Counterpart of Fig. \ref{fig:cyl2d_1_20_pdf} but for the experimental circular cylinder wake flows (Re = 2600), CR20 database.}
    \label{fig:vki_pdf_5}
\end{figure}

Figure \ref{fig:vki_pdf_5} shows the normalized error probability distribution function for the streamwise and normal velocity components across the reconstructions obtained from the 1:20 low-resolution dataset, using both the SVD-SR and HOSVD-SR approaches. Both methods exhibit an asymmetrical distribution centred around $0$ on the error axis for the normal velocity component. However, the normalized error for the streamwise velocity component resembles a right-skewed distribution, with peak values around $8$ $\times$ $10^{-2}$ for the SVD-based approach and $4$ $\times$ $10^{-2}$ for the HOSVD-based approach. The probability that the normalized error is zero for the SVD-SR method is around $32\%$ for the normal velocity component and $30\%$ for the streamwise velocity component, while for the HOSVD-SR are $41\%$ and $37\%$ respectively. 

The similitude between the reconstruction RRMSE obtained with the HOSVD-SR and SVD-SR approaches is due to the fact that both methods retain a similar number of dominant modes. As a consequence, the reconstruction captures primarily the dynamics associated with the large-scale flow structures, while small-scale features are filtered out. This limitation leads to a comparable level of error in both reconstructions. However, as observed in the error distribution and confirmed by the computed RRMSE values, the HOSVD-SR approach consistently outperforms SVD-SR, resulting in lower reconstruction errors.

\begin{figure}[H]
    \centering
    \includegraphics[width=14 cm ]{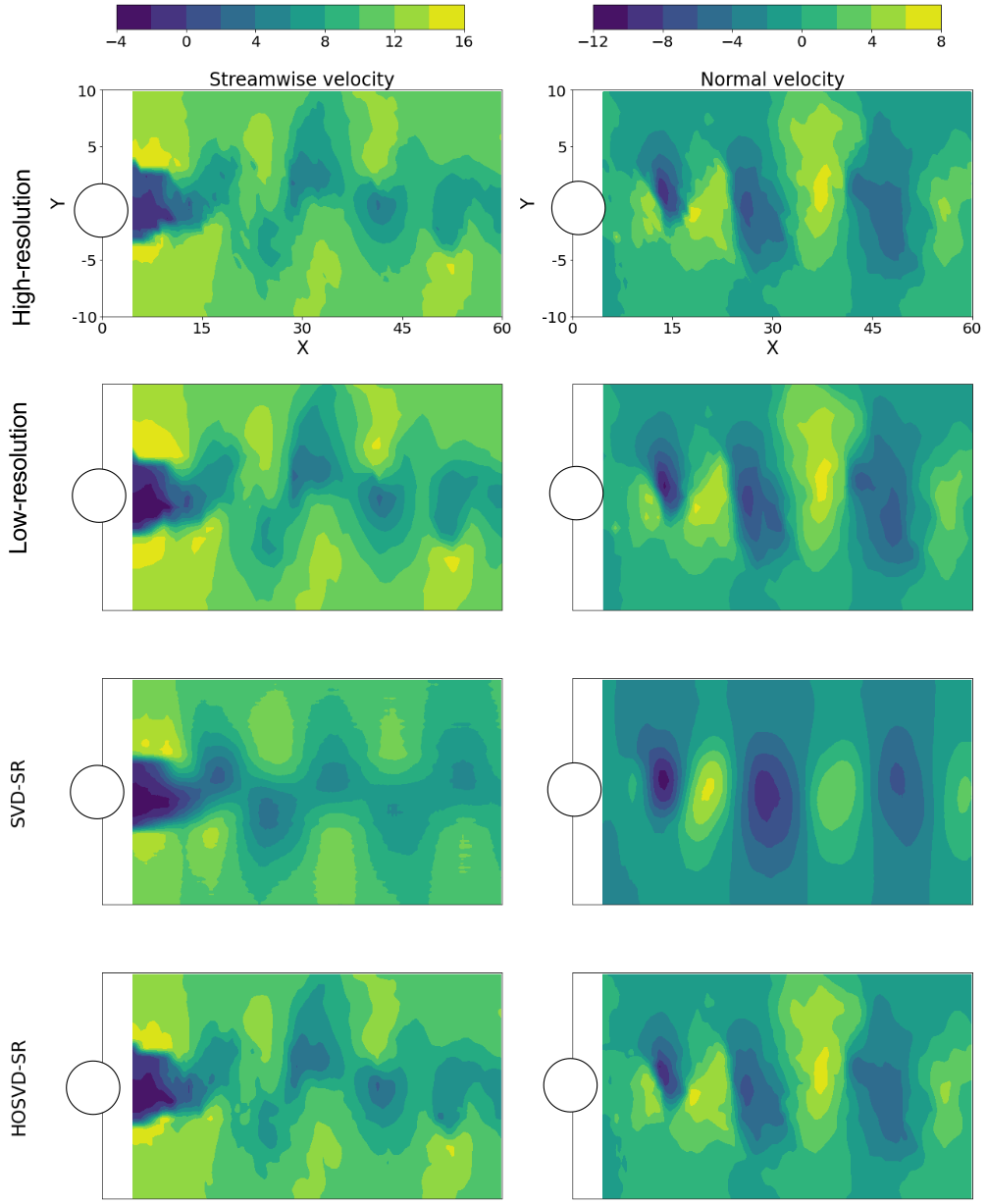} 
    \caption{Same as Fig. \ref{fig:cyl2d_1_40} but for the experimental circular cylinder wake flows (Re = 2600), CR5 database.}
    \label{fig:22_VKI}
\end{figure}

Figure \ref{fig:22_VKI} presents the resulting reconstructions after applying the two resolution enhancement approaches to the 1:5 compression ratio dataset with $20$ SVD modes on each $x$ and $y$ spatial dimensions. As previously mentioned, with this compression rate, the reconstruction RRMSE of the SVD-SR method is more than twice larger that of the HOSVD approach for both the streamwise and normal velocity components. Although the RRMSE values for the SVD-SR are relatively low, it yields a quasi-symmetrical shape in the wake region and fails to precisely reconstruct small and irregular flow structures detached from the bluff body. In contrast, the proposed HOSVD-based approach accurately reconstructs these features, maintaining the detail and distribution of the flow along the wake. This is due to the ability of HOSVD to independently extract the dominant patterns as SVD mode matrices related to each one of the dimensions of the database, preserving the small scales.

\begin{figure}[H]
    \centering
    \includegraphics[width=\textwidth]{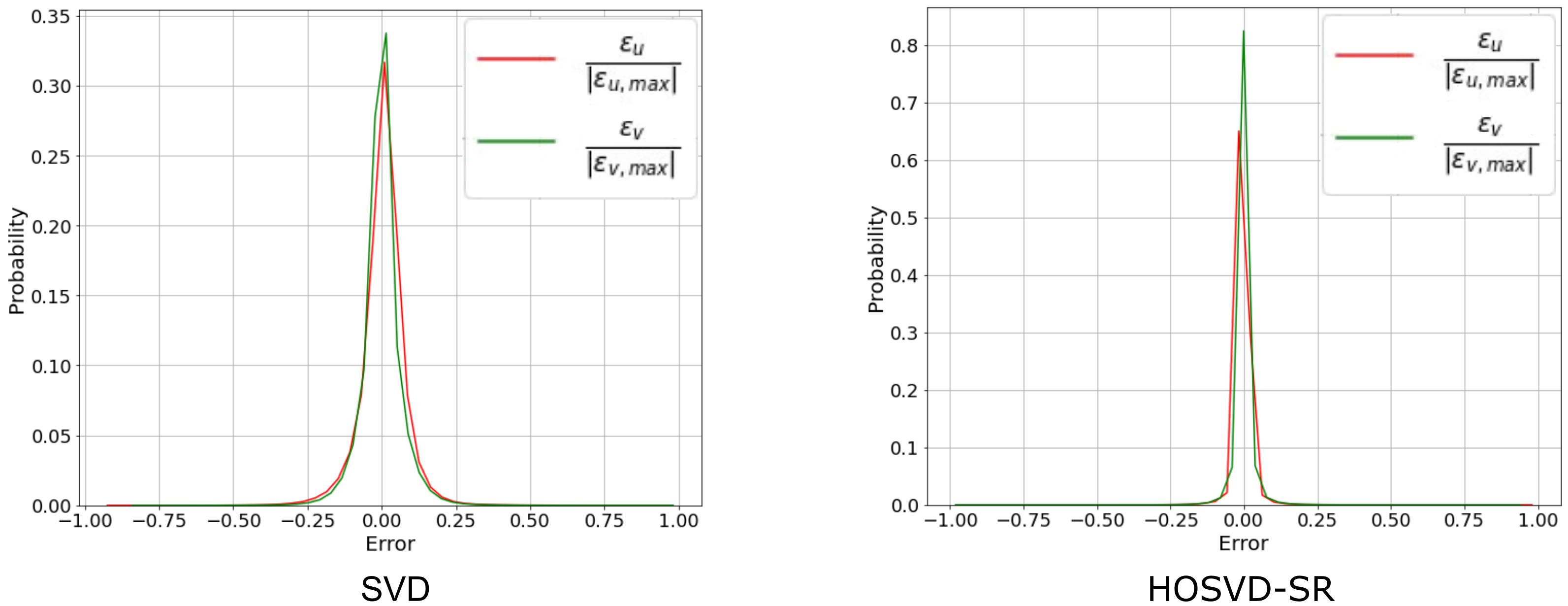} 
    \caption{Counterpart of Fig. \ref{fig:cyl2d_1_40_pdf} but for the experimental circular cylinder wake flows (Re = 2600), CR5 database.}
    \label{fig:vki_pdf_22}
\end{figure}

The probability distribution functions for the 1:5 low resolution database using both the SVD-SR and HOSVD-SR approaches are shown in Fig. \ref{fig:vki_pdf_22}. The SVD-based method presents a lean, asymmetric distribution with peaks at 0 on the error axis, exhibiting occurrence probabilities for the normalized error being equal to zero of 37\% and 31\% for the normal and streamwise velocity components, respectively. In contrast, the HOSVD-based approach has an asymmetrical distribution with an 82\% probability for the normalized error to be 0 in the normal velocity component and a peak at 0.03 with a 65\% probability of occurrence for the streamwise velocity component. These results indicate that the HOSVD-based methodology achieves lower reconstruction errors compared to the SVD-SR approach, with the extent of improvement depending on the compression ratio and the number of SVD modes retained after applying HOSVD or SVD.

As previously mentioned, the novel HOSVD-based resolution enhancement methodology proposed is highly dependent on the number of SVD modes available and retained. From an efficiency standpoint, the goal is to achieve simultaneously the most accurate and clean reconstruction with the fewest retained SVD modes and the lowest resolution data possible. In order to evaluate the influence of the number of retained SVD modes, various tests were conducted on multiple low resolution databases with different numbers of retained SVD modes for each spatial dimension.

Table \ref{tab:vki_modes} presents the reconstruction RRMSE for various combinations of compression rates and retained SVD modes. The results indicate that, for each compression rate, the reconstruction RRMSE decreases as the number of retained SVD modes increases. It is also observed that the reconstruction RRMSE is higher for reconstructions obtained from the 1:2 and 1:5 low resolution databases with fewer retained SVD modes. This outcome is related to how the available data is organized and stored in the mode matrices. Lower compression rates (e.g., 1:2, 1:5) contain more information, including local minima and maxima, which can be captured by the principal SVD modes. If fewer modes are retained, some critical data may be lost, leading to higher reconstruction errors.

\begin{figure}[H]
    \centering
    \includegraphics[height=17.5 cm]{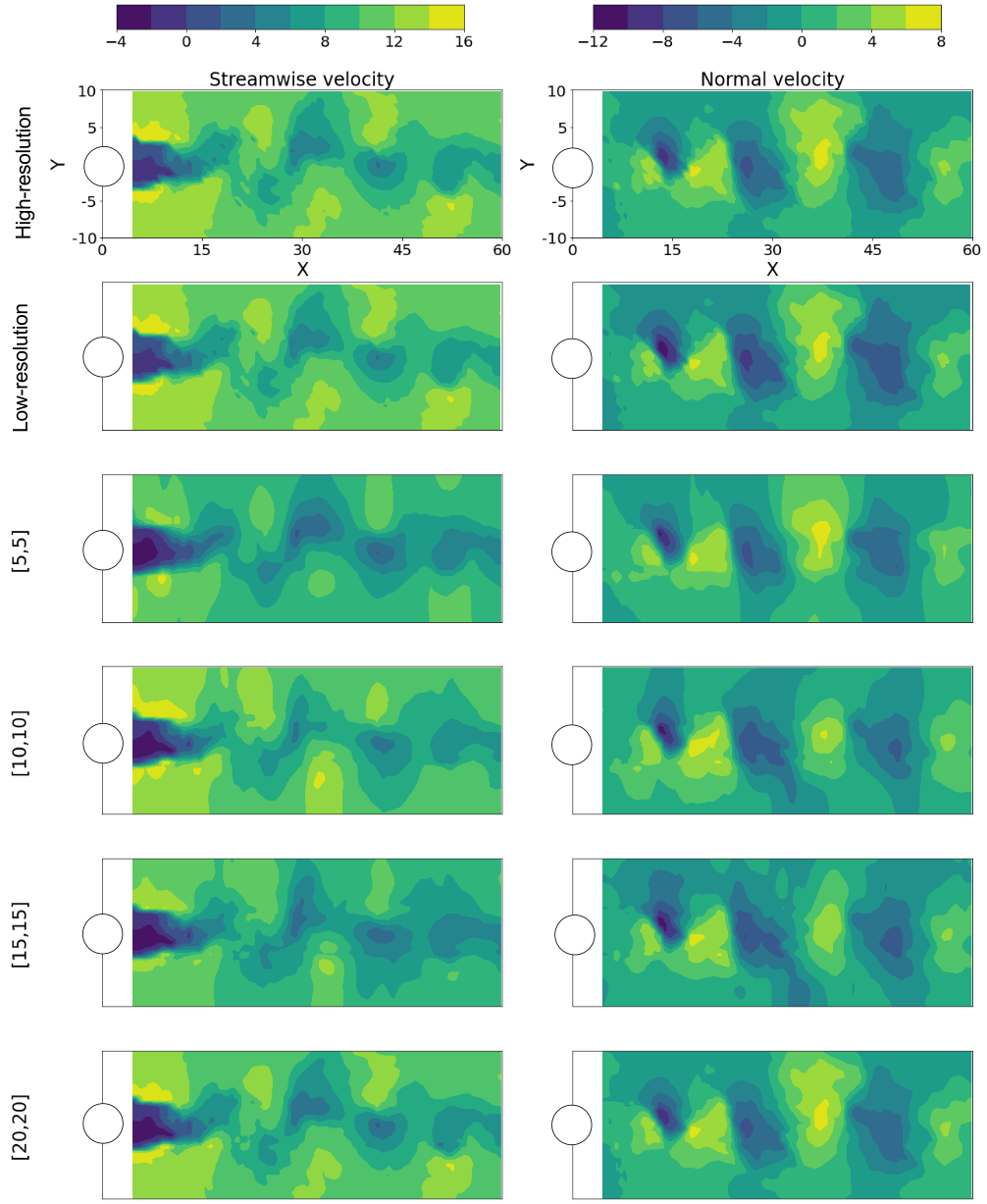}
    \caption{Experimental database of a turbulent cylinder wake flows with Re = 2600, CR20: Comparison between  the HOSVD-SR approach for a compression ratio of 1:5 database with 5, 10, 15 and 20 SVD modes retained along each spatial dimension.}
    \label{fig:vki_modes_22}
\end{figure}

Figure \ref{fig:vki_modes_22} illustrates the reconstructions obtained using the proposed HOSVD super-resolution method with different numbers of retained SVD modes for a database with a compression ratio of 1:5. The reconstruction of small flow structures improves as the number of retained SVD modes increases. For instance, with 5 retained modes on each spatial dimension, the reconstruction has an RRMSE of 4.05 $\times$ $10^{-2}$ and successfully recreates the original flow behavior with regular and smoothed shapes, which reflects the de-noising capabilities of HOSVD. When 20 spatial SVD modes are retained on each dimension, which is close to the maximum available modes, the RRMSE decreases to 1.17 $\times$ $10^{-2}$, being the small scale structures precisely reconstructed.

\begin{table}[H]
\caption{Same as Tab. \ref{tab:2dcyl_modes} but for the experimental database of turbulent cylinder wake flows with Re = 2600.}
\centering
\begin{tabular}{lcccc}
\toprule
\multicolumn{5}{c}{\textbf{RRMSE} $\times 10^{-2}$} \\
\midrule
\multirow{2}{*}{\textbf{compression rate}} & \multicolumn{4}{c}{\textbf{Number of Retained Modes}} \\
          & 5 & 10 & 15 & 20 \\
\midrule
1:2       & 4.05 & 2.06 & 1.49 & 1.17 \\
1:5       & 4.07 & 3.04 & 2.93 & - \\
1:10      & 4.16 & 3.21 & - & - \\
1:20      & 4.55 & - & - & - \\
\bottomrule
\end{tabular}

\label{tab:vki_modes}
\end{table}

\section{Conclusions}\label{conclusions}

This article introduces HOSVD-SR which is a novel hybrid machine-learning ROM developed to enhance the spatial resolution of fluid dynamics databases generated through numerical or experimental approaches. The method combines modal decomposition with deep learning strategies. It takes as input a low-resolution database in tensor form, which is decomposed into SVD mode matrices along each dimension. The mode matrices corresponding to the spatial dimensions are then expanded to high resolution using a fully connected neural network. This methodology has been validated in both numerical and experimental databases, in laminar and turbulent flow regimes. 

The proposed HOSVD-SR methodology effectively enhances the spatial resolution of fluid dynamics data by combining HOSVD with Deep Learning. This modal decomposition technique efficiently extracts the underlying physics of the fluid dynamics phenomena within the database, while the implemented neural network addresses the non linearity.

The HOSVD-SR approach outperforms the SVD method across the tested datasets and various compression rates, achieving up to $30\%$ lower RRMSE values. This is attributed to HOSVD's ability to identify the singular values associated with each component of a fluid dynamics database, efficiently filtering noise and reducing its dimensionality, enhancing the reconstruction accuracy and robustness. 

The HOSVD-SR method consistently achieves RRMSE values below $10\%$ across various benchmark cases, compression rates, and numbers of retained SVD modes. Its robustness and high accuracy highlight the value of the proposed methodology for analysing complex fluid dynamics phenomena, such as turbulent flows, where high-resolution data is essential for accurate representation and analysis.

The number of spatial SVD modes available is closely related to the compression rate; lower compression ratio datasets allow to retain a greater number of SVD modes. It is worth mentioning that a higher number of retained modes may require more computational resources, but the accuracy of the reconstruction also increases, specially for the small scales.

The proposed HOSVD-SR methodology has demonstrated high accuracy in reconstructing fluid flow databases with two and three spatial dimensions at various Reynolds numbers. Its effectiveness and robustness make it a powerful tool for fluid dynamics analysis. Additionally, the selection of spatial SVD modes provides flexibility, allowing users to balance computational efficiency and reconstruction fidelity. In some cases, using fewer modes can help eliminate noise, reduce spatial redundancies, and improve numerical stability. Future work may explore alternative neural network architectures to further enhance the reconstruction of small-scale structures when required.

\section{Acknowledgements}

The authors acknowledge the MODELAIR project that has received funding from the European Union’s Horizon Europe research and innovation programme under the Marie Sklodowska-Curie grant agreement No. 101072559. S.L.C. acknowledges the ENCODING project that has received funding from the European Union’s Horizon Europe research and innovation programme under the Marie Sklodowska-Curie grant agreement No. 101072779. The results of this publication reflect only the author's view and do not necessarily reflect those of the European Union. The European Union can not be held responsible for them. The authors acknowledge the grant PLEC2022-009235 funded by MCIN/AEI/ 10.13039/501100011033 and by the European Union “NextGenerationEU”/PRTR and the grant PID2023-147790OB-I00 funded by MCIU/AEI/10.13039 /501100011033 /FEDER, UE. The authors gratefully acknowledge the Universidad Politécnica de Madrid (www.upm.es) for providing computing resources on the Magerit Supercomputer.

\appendix
\renewcommand{\thesection}{\Alph{section}} 
\renewcommand{\thetable}{\Alph{section}.\arabic{table}} 

\section{APPENDIX}\label{annex:anexo}

This section shows in detail the hyperparemeters used to train the neural networks for both the SVD-SR and HOSVD-SR approaches with the highest compression rations on each database tested.

\begin{table}[H]
\centering
\caption{HOSVD-SR hyperparameters applied to the experimental circular cylinder wake database (CR 1:20).}
\label{tab:hyperparameters_3d}
\begin{tabular}{lll}
\toprule
\textbf{Hyperparameter} & $\textbf{W}^1$ & $\textbf{W}^2$ \\
\midrule
neurons (Dense 1 – ``ReLu'')  & 64 & 64  \\
neurons (Dense 2 – ``ReLu'') & 128 & 128 \\
neurons (Dense 3 – ``linear'')   & 995   &  5388 \\
Batch size & 8 & 8 \\
Loss Function (training) & MSE & MSE \\
Loss Function (validation) & MAE & MAE \\
Learning rate & 0.0005 & 0.0005 \\
\bottomrule
\end{tabular}
\end{table}

\begin{table}[H]
\centering
\caption{HOSVD-SR hyperparameters applied to the experimental circular cylinder wake database (CR 1:20).}
\label{tab:hyperparameters_3d}
\begin{tabular}{lll}
\toprule
\textbf{Hyperparameter} & $\textbf{W}^1$ & $\textbf{W}^2$ \\
\midrule
neurons (Dense 1 – ``ReLu'')  & 64  & 64  \\
neurons (Dense 2 – ``ReLu'') & 96 & 128 \\
neurons (Dense 3 – ``linear'')  & 555   & 4515  \\
Batch size & 4 & 4 \\
Loss Function (training) & MSE & MSE \\
Loss Function (validation) & MAE & MAE \\
Learning rate & 0.0005 & 0.0005 \\
\bottomrule
\end{tabular}
\end{table}

\begin{table}[H]
\centering
\caption{HOSVD-SR hyperparameters applied to the 3D circular cylinder wake database (CR 1:20).}
\label{tab:hyperparameters_vki}
\begin{tabular}{llll}
\toprule
\textbf{Hyperparameter} & $\textbf{W}^1$ & $\textbf{W}^2$ & $\textbf{W}^3$ \\
\midrule
neurons (Dense 1 – ``ReLu'') & 64  & 64  & 64 \\
neurons (Dense 2 – ``ReLu'') & 128 & 96 & 96\\
neurons (Dense 3 – ``linear'') & 500   & 80 & 192  \\
Batch size & 16 & 16 \\
Loss Function (training) & MSE & MSE & MSE \\
Loss Function (validation) & MAE & MAE & MAE \\
Learning rate & 0.0005 & 0.0005 & 0.0005 \\
\bottomrule
\end{tabular}
\end{table}

\begin{table}[H]
\centering
\caption{SVD-SR hyperparameters applied to the 2D circular cylinder wake database (CR 1:20).}
\label{tab:hyperparameters_3d}
\begin{tabular}{lll}
\toprule
\textbf{Hyperparameter} & $\textbf{U}$ & $\textbf{V}$ \\
\midrule
neurons (Dense 1 – ``ReLu'')  & 10 & 10 \\
neurons (Dense 2 – ``linear'')  & 995   &  5388 \\
Batch size & 16 \\
Loss Function (training) & MSE & MSE \\
Loss Function (validation) & MAE & MAE \\
Learning rate & 0.001 & 0.001 \\
\bottomrule
\end{tabular}
\end{table}

\begin{table}[H]
\centering
\caption{SVD-SR hyperparameters applied to the 3D circular cylinder wake database (CR 1:20).}
\label{tab:hyperparameters_SVD_vki}
\begin{tabular}{lll}
\toprule
\textbf{Hyperparameter} & \textbf{U} & \textbf{V}\\
\midrule
neurons (Dense 1 – ``ReLu'') & 128  & 128 \\
neurons (Dense 2 – ``linear'') & 500  & 15360 \\
Batch size & 16 & 16 \\
Loss Function (training) & MSE  & MSE \\
Loss Function (validation) & MSE  & MSE \\
Learning rate & 0.001   & 0.001\\
\bottomrule
\end{tabular}
\end{table}

\begin{table}[H]
\centering
\caption{SVD-SR hyperparameters applied to the experimental circular cylinder wake database (CR 1:20).}
\label{tab:hyperparameters_SVD_vki}
\begin{tabular}{lll}
\toprule
\textbf{Hyperparameter} & \textbf{U} & \textbf{V}\\
\midrule
neurons (Dense 1 – ``ReLu'') & 128  & 128 \\
neurons (Dense 2 – ``linear'') & 555   & 4515  \\
Batch size & 16 \\
Loss Function (training) & MSE  & MSE \\
Loss Function (validation) & MSE  & MSE \\
Learning rate & 0.001   & 0.001 \\
\bottomrule
\end{tabular}
\end{table}

\end{document}